\documentclass{ar-1col-S2O}
\usepackage[numbers]{natbib}
\usepackage{url}
\usepackage{xcolor}

\usepackage{dsfont}
\usepackage{amsmath}
\usepackage{amssymb}
\usepackage{cleveref}
\usepackage{nicefrac}
\usepackage{placeins}
\usepackage{subcaption}
\usepackage{xfrac}

\jname{submitted to: Annual Review of Nuclear and Particle Science, Volume 76}
\jvol{AA}
\jyear{2026}
\doi{10.1146/annurev-nucl-102422-040841}

\newcommand{\N}[1]{\ensuremath{\tilde{\chi}_{#1}^0}}
\newcommand{\C}[1]{\ensuremath{\tilde{\chi}_{#1}^{\pm}}}
\newcommand{\slepton}[1][]{\ensuremath{\tilde{\ell}_{#1}}}
\newcommand{\selectron}[1][]{\ensuremath{\tilde{e}_{#1}}}
\newcommand{\smuon}[1][]{\ensuremath{\tilde{\mu}_{#1}}}
\newcommand{\stau}[1][]{\ensuremath{\tilde{\tau}_{#1}}}

\newcommand{\deltaMpm}[1][]{\ensuremath{\Delta m_{\pm}^{#1}}}
\newcommand{\deltaMo}[1][]{\ensuremath{\Delta m_{0}^{#1}}}

\newcommand{\met}{\ensuremath{p_T^{\text{miss}}}}
\newcommand{\me}{\ensuremath{p^{\text{miss}}}}

\newcommand{\axe}{\ensuremath{A\times\varepsilon}}

\begin{document}

\markboth{L. Jeanty and L. Lee}{Rare and Experimentally Challenging Supersymmetry Signatures}

\title{Rare and Experimentally Challenging Supersymmetry Signatures}

\author{Laura Jeanty$^1$ and Lawrence Lee$^2$
\affil{$^1$Department of Physics, Institute for Fundamental Science, University of Oregon, Eugene, OR, USA, 97403; email: ljeanty@uoregon.edu}
\affil{$^2$Department of Physics \& Astronomy, University of Tennessee, Knoxville, Knoxville, TN, USA, 37996; email: llee@utk.edu}}

\begin{abstract}
Supersymmetry has long played a central role in the search for physics beyond the Standard Model at colliders, providing a comprehensive and internally consistent framework for generating well-motivated experimental signatures. For more than fifteen years of LHC operation, the CMS and ATLAS collaborations have achieved remarkable sensitivity to a wide range of supersymmetric signatures. Despite this unprecedented reach, no conclusive evidence for supersymmetry has emerged. If supersymmetry is nature's solution to outstanding questions in particle physics, it is necessarily challenging to find. In this article, we review supersymmetric signatures that are particularly rare or otherwise challenging, with a focus on searches at the Large Hadron Collider. We highlight \textit{experimental} challenges relating to detector constraints and analysis difficulties, in addition to \textit{model} challenges in the interpretation and optimization of searches. We also identify regions of signature space that remain comparatively unconstrained and therefore represent promising targets for future exploration.
\end{abstract}

\maketitle

\tableofcontents

\section{INTRODUCTION}

Understanding electroweak physics has been a central scientific goal for over a half century. In the mid-1960s, the theoretical unification of the electromagnetic and weak interactions laid the groundwork for the modern Standard Model (SM) of particle physics. The necessity of electroweak symmetry breaking was indicated by the masses of the $W^{\pm}$ and $Z^0$ bosons, which were correctly predicted to be $\mathcal{O}(100)$~GeV decades before their discovery in 1983. The simplest mechanism for breaking electroweak symmetry was the Higgs mechanism, proposed in 1964~\cite{Higgs:1964pj, Englert:1964et,Guralnik:1964eu}. Nearly 50 years later, the ATLAS and CMS collaborations at the Large Hadron Collider (LHC) announced discovery of the Higgs Boson, the particle that emerges from the Higgs mechanism~\cite{ATLAS:2012yve, CMS:2012qbp}. With the discovery of the Higgs Boson, electroweak symmetry breaking is both mathematically described and experimentally confirmed.

Important questions remain unanswered. What is the shape of the Higgs potential? Why is electroweak symmetry broken at a scale of 246 GeV, leading to a Higgs mass of a similar order at 125 GeV? The former must be experimentally measured. The latter has deep implications for the philosophical validity of the Standard Model in nearly any extension that includes physics at higher energy scales. Quantum corrections to the square of the measured Higgs mass arise from interactions with virtual particles via loop diagrams; these corrections are quadratically proportional to both the mass of the interacting particle and the energy scale at which new physics appears. Given that quantum gravitational effects are important at the Planck scale, $\mathcal{O}(10^{18})$~GeV, and assuming no new physics between the electroweak and Planck scales, the natural expectation for the squared Higgs Boson mass is roughly $10^{36}$~GeV$^2$ instead of the observed $10^{4}$~GeV$^2$. This requires a fine-tuning between the bare Higgs mass squared and its quantum corrections at the level of one part in $10^{32}$, an exquisite balance at extreme odds with the expectation that our universe is not unnaturally atypical~\cite{primer,Gildener:1976ai,Susskind:1978ms}.

\textit{Supersymmetry} (SUSY) is a theoretical framework for extending the Standard Model that was first proposed in the early 1970s~\cite{Wess:1974tw}, expanding the spacetime symmetries of the Poincar\'{e} group to include anti-commuting operators that would correspond to a new internal fermion-boson symmetry. Each boson (fermion) would sit in a supermultiplet with its new fermionic (bosonic) partner \emph{sparticle}. SUSY seemed to violate the earlier Coleman-Mandula no-go theorem which argued that spacetime symmetries and internal symmetries must factorize trivially~\cite{Coleman:1967ad}, until it was shown by the Haag–Łopuszański–Sohnius theorem to be the \emph{only} such exception possible~\cite{Haag:1974qh}. Within a few years of its introduction, SUSY was found to fit naturally into string theory, helping to solve some crucial problems in early string theory such as the inability to accommodate fermions. Such arguments in field theory and model building suggest that SUSY might exist at \emph{some} energy scale.

In the 1970s, the \textit{Higgs Naturalness Problem} discussed above, induced by the \textit{Gauge Hierarchy Problem}, was identified~\cite{Wilson:1970ag, Susskind:1978ms}, and it was quickly pointed out that an electroweak-scale SUSY could provide the mechanism for stabilizing the Higgs mass, removing its quadratic UV sensitivity~\cite{Dimopoulos:1981zb, Witten:1981nf}. The natural cancellations that appear in the Higgs mass corrections due to SUSY lead to the stability of the observed Higgs mass without requiring parameters that are fine-tuned to many orders of magnitude.

Beyond stabilizing the Higgs mass, supersymmetry appearing near the electroweak scale could answer a number of questions on which the SM is silent. If the lightest supersymmetric particle (LSP) is stable and electrically neutral, it is a prime dark matter (DM) candidate~\cite{Goldberg:1983nd,Ellis:1983ew}. If such an LSP has a mass near the TeV scale, it would be a weakly interacting massive particle (WIMP) DM candidate whose cosmological abundance could match the observed thermal relic density. Moreover, the contributions from superpartners allow for the unification of Standard Model gauge couplings at high energy, paving the way to unification of the gauge forces. Additionally, the SM does not contain enough CP violation to explain the observed matter-antimatter asymmetry in the universe, but with the addition of a plethora of SUSY couplings comes new opportunities for CP-violating phases, more than enough to accommodate the observed asymmetry~\cite{Dine:2003ax}. For a theoretical and experimental introduction to SUSY, see Refs.~\cite{primer} and~\cite{ATLAS:2024lda}, respectively.

The search for electroweak-scale SUSY has been ongoing since its theoretical inception in the early 1980s. In fact, precisely-electroweak-scale SUSY has been ruled out by experiment, and near-electroweak-scale SUSY instead is the target of modern experiments~\cite{Baer:2025zqt}. A simple reading of the naturalness solution predicted by SUSY models might suggest that TeV-scale SUSY is strictly less desirable than a sparticle spectrum at $\mathcal{O}(100)$~GeV, but naive dimensional analysis~\cite{Manohar:1983md} has led to successful predictions for particles appearing around an order of magnitude above a known scale, as in the case of the prediction of the $\rho$ meson mass, an order of magnitude above the pion decay scale, or the charm quark mass, an order of magnitude above the QCD scale in the GIM mechanism. With this naturalness heuristic, if nature repeated itself with the electroweak scale, we would naively expect Beyond the Standard Model (BSM) physics not \emph{at} $\mathcal{O}(100)$~GeV, but rather at $\mathcal{O}(1)$~TeV.

\begin{figure}
\includegraphics[width=4in]{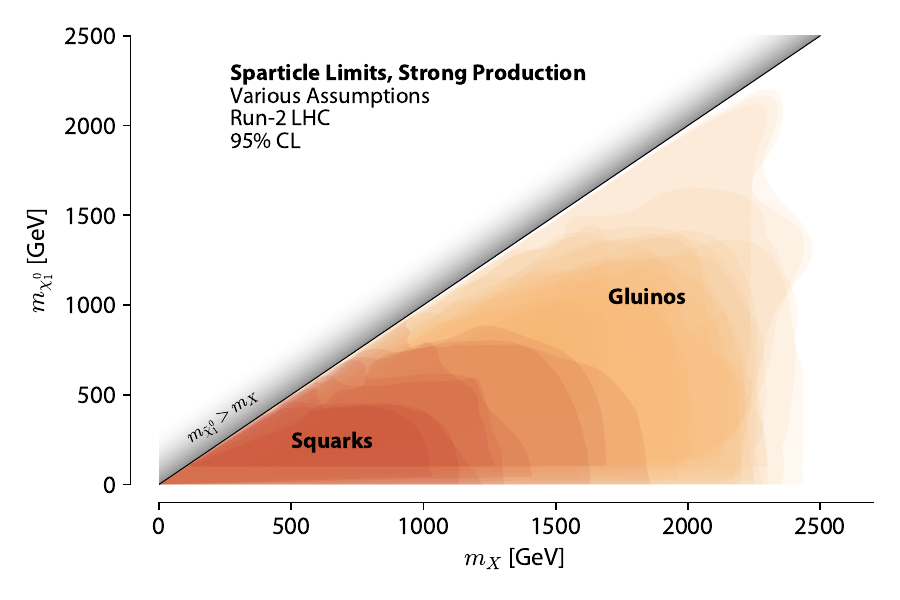}\\
\includegraphics[width=4in]{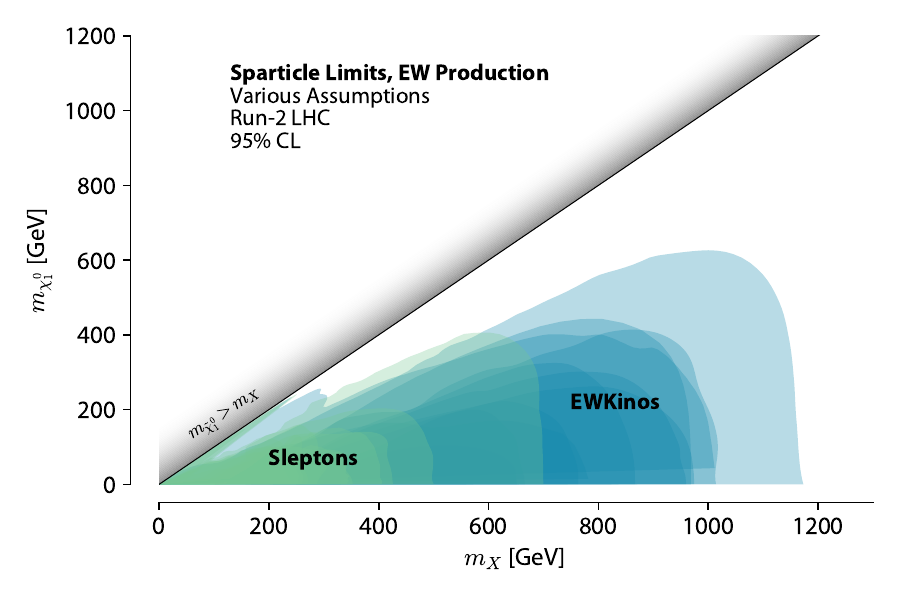}
\caption{A wide array of conventional SUSY search results from the ATLAS and CMS experiments is shown for strongly produced sparticles (top) and those produced via EW processes (bottom), as a function of the mass of the produced sparticle, $m_X$, and the mass of the neutralino LSP, $m_{\N1}$. Different colors illustrate searches with varying assumptions about the sparticle couplings and decays~\cite{ATLAS:2024lda,Sekmen:2025bxv,ATL-PHYS-PUB-2024-014,CMSSUSYSummaryPlots}. 
}
\label{fig:vanilla_susy}
\end{figure}

In recent memory, there have been extensive search programs at S$p\bar{p}$S, LEP~\cite{LEP_susy}, the Tevatron~\cite{Carena:1997mb,ParticleDataGroup:2024cfk,Sopczak:2014aua}, $B$-factories~\cite{Cohen:1996sq}, and now at the LHC~\cite{ParticleDataGroup:2024cfk,theoristSUSY_review,ATLAS:2024lda,Sekmen:2025bxv,ATL-PHYS-PUB-2024-014,CMSSUSYSummaryPlots}. Thus far, searches for SUSY have produced null results. These results provide extensive exclusion power, helping to guide further searches and providing important constraints on SUSY models. Figure~\ref{fig:vanilla_susy} shows a large collection of overlapping exclusion contours for traditional SUSY searches from the LHC experiments, showing robust coverage across a wide range of simplified-model parameters. Despite these exclusions, in this review, we will argue that the possibility of TeV-scale SUSY is still compelling, like the chance of rain on a dry but still cloudy afternoon. With the existing experimental constraints, however, such a SUSY must be \textit{rare} to produce, \textit{challenging} to find, or both.

In the following, we summarize the state of searches for rare and challenging supersymmetry signatures, with a focus on recent results from the LHC, discuss what makes a signature experimentally challenging, summarize the evolution of the program, and identify areas of interest for the rest of the LHC era, the HL-LHC, and future colliders. Although we focus on rare and challenging \emph{signatures}, we organize the discussion by the SUSY considerations that motivate those signatures; SUSY models provide a systematic and comprehensive framework for understanding the signatures in context. This work is not intended to be an exhaustive review of searches for supersymmetry at the LHC. For comprehensive summaries of the ATLAS and CMS search programs, please see Refs.~\cite{ATLAS:2024lda} and~\cite{Sekmen:2025bxv}, respectively. For a recent review of the theoretical implications of the results of these search programs, please see Ref.~\cite{theoristSUSY_review}.

\section{RARE \& CHALLENGING SIGNATURES}

The experimental difficulty in discovering a new phenomenon is, of course, relative to the tools at hand. After the discovery of the positron, it took over 20 years of dedicated accelerator development to produce and detect the anti-proton. By the early 1980s, the production and accumulation of anti-protons became sufficiently efficient that they themselves were a main ingredient in the $Sp\bar{p}S$ collider. Discovery is also highly sensitive to small changes in parameters: had the Higgs boson's mass differed by 30\% in either direction, it might have been observed 5-20 years earlier by either the Tevatron or LEP.

As long as we experimentalists are doing a reasonable job, what we haven't yet discovered at any given time is \textit{by definition} rare and/or challenging, as long as we're looking in the right places. Note that the ``and/or'' is essential; there is a subtle interplay between \emph{rare} and \emph{challenging}.

\begin{marginnote}[]
\entry{Simplified Model}{A framework in which a small number of sparticles and one production and decay process are considered while all others are assumed to be decoupled, at a mass scale too heavy to affect the considered production and decay. The simplified model approach was adopted as the standard during Run-1 of the LHC to ease external interpretation of results. Scans are performed over experimentally accessible EW-scale masses instead of Lagrangian parameters.}
\end{marginnote}

A rare \emph{process} can be defined by its production cross-section and relevant branching ratio, $\sigma\times BR$, relative to the integrated luminosity, $\mathcal{L}$, of the available dataset. Since any given sparticle production cross-section in a simplified model drops rapidly with mass, we must also specify the mass scale at which we define rare. A rare \emph{signature} can also include effects related to detector acceptance, $A$. As a useful metric for quantifying rare signatures, we consider that using a $CL_s$ prescription, a typical Poisson counting search with negligible background can exclude, at the 95\% confidence level (CL), a signal corresponding to approximately three expected events~\cite{ParticleDataGroup:2024cfk}. In principle a discovery can be made with fewer than 3 observed events, given a zero-background SM expectation; for example, the discovery of the $\Omega^-$ baryon was made with a single event.

Rare processes are intrinsically challenging to observe, as collecting, reconstructing, and analyzing quality data is not an automatic process. Accumulating a large dataset requires years of smooth accelerator and detector operation in the face of aging hardware, radiation damage, and personnel and funding challenges. Acquiring a dataset that increases more-than-linearly with time is even more challenging, requiring accelerator, detector, trigger, and data-acquisition improvements in addition to routine maintenance and operation tasks. Storing and processing a large dataset presents significant software and \textbf{computing constraint} challenges. At the LHC, many of these challenges are shared among the entire measurement and search program. A rare signature that does not require special trigger or reconstruction techniques and without significant SM backgrounds might not pose additional analysis-level challenges, such as a high $p_T$, multi-lepton signature from the decay of a very heavy sparticle. Nonetheless, it is always a challenge to fulfill the experimenter's responsibility to optimize the search for each signature as much as possible. 

\begin{figure}
\includegraphics[width=5in]{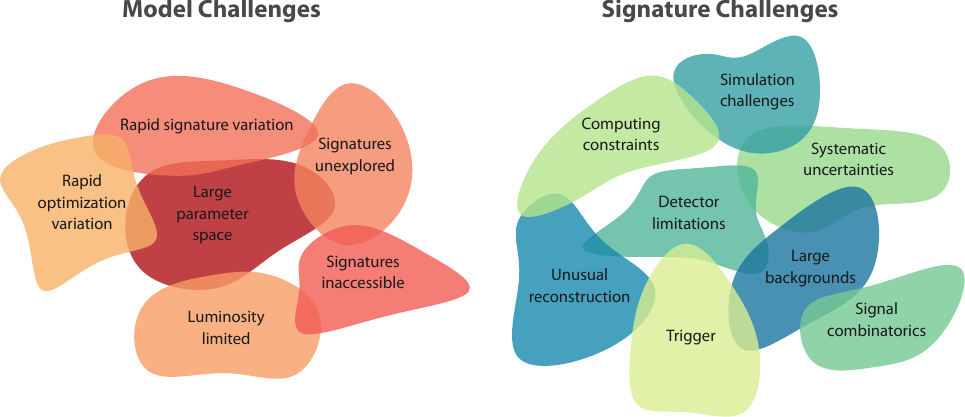}
\caption{An abstract representation of the challenges faced by modern searches for supersymmetry. The challenges inherent to searches for a particular SUSY model are distinct from the challenges inherent to a particular experimental signature. The overlaps between challenges are intended to be illustrative of actual overlapping challenges, but they are neither exhaustive nor prescriptive.}
\label{fig:challenges}
\end{figure}

While there is no universal definition of \textit{challenging} in the context of searches for SUSY, in this work, we separate out two different realms of challenges: those that are intrinsic to the model, and those that are instead intrinsic to a particular signature. A specific SUSY model or corner of SUSY parameter space might be challenging to probe because it has a large parameter space in which both signatures and the optimizations for those signatures vary quickly with model parameters. It may predict signatures that are either inaccessible or unexplored at a particular experiment, and if it favors a certain mass scale, the accessibility of signatures may be limited by $\mathcal{L}$. The challenges inherent to experimental signatures sometimes (but not always) affect the efficiency $\varepsilon$ of reconstructing, identifying, and selecting the signature. More broadly, signature challenges include detector limitations, unusual reconstruction requirements, computing constraints, simulation challenges, trigger limitations, large backgrounds, systematic uncertainties, and issues arising from signal object combinatorics. These challenges are illustrated graphically in Figure~\ref{fig:challenges} and will be discussed throughout this review.

One example that will resurface several times in this review is the case of Long-Lived Particle (LLP) signatures at colliders. In many SUSY models, BSM particles gain measurably macroscopic lifetimes. Loosening the assumption that heavy BSM particles promptly decay quickly creates experimental challenges as the detectors, trigger systems, and reconstruction algorithms are largely built to measure SM particles traveling projectively from the collision point. Efficient and thorough coverage of SUSY LLP signatures requires dedicated \textbf{unusual reconstruction} efforts to achieve good sensitivity throughout the lifetime range. For a complete review of the challenges of LLP searches, please see Refs~\cite{Lee:2018pag, Jeanty:2025wai}.

\section{MSSM-INSPIRED SIGNALS}

The Minimal Supersymmetric Standard Model (MSSM) is a framework that preserves the structure of the SM while adding only the minimal number of new particles and terms in the Lagrangian that are needed for a self-consistent supersymmetric extension of the SM. It includes parameters that explicitly break supersymmetry and allows for different masses between sparticles and their SM counterparts, but it doesn't specify the origin of these terms~\cite{Haber:1984rc,primer}. As pointed out in Ref.~\cite{primer}, all the particles in the MSSM that must be massless before electroweak symmetry breaking, and therefore those that could reasonably be expected to be light, have already been discovered (\textit{i.e.} the SM particles). All sparticles can be massive without breaking EW symmetry, and therefore it might be reasonable to expect that they exist at a different mass scale from their SM partners.

The superpartners to the gluon and quarks (the gluino $\tilde{g}$ and squarks $\tilde{q}$) would be produced copiously at hadron colliders via strong interactions. Searches for the superpartner to the top quark (the stop $\tilde{t}$) have played a special role in the naturalness question, as the heaviness of the top quark plays a out-sized role in the naturalness problem~\cite{primer, theoristSUSY_review}. For a recent review of searches for stops, see Ref.~\cite{Franceschini:2023nlp}. The discovery of the Higgs Boson with a mass of 125 GeV places significant constraints on the stop sector within the MSSM; in the generic case, the mass of the Higgs requires either large mixing between the superpartners of the left-handed and right-handed tops, $\tilde{t}_L$ and $\tilde{t}_R$, and $m_{\tilde{t}} \gtrsim 1$~TeV, or significantly heavy stops with $m_{\tilde{t}} \gtrsim 5$~TeV~\cite{Draper:2011aa}.

While the particle content in the MSSM is independent of a specific SUSY-breaking mechanism, an expectation for sparticle masses and more commonly the ordering of these masses often emerges from specific SUSY-breaking models. For a detailed discussion of the phenomenological consequences of different SUSY-breaking mechanisms, see Ref.~\cite{primer}. For a recent discussion of naturalness, see Ref.~\cite{Baer:2025zqt}. In the following sections, we follow the simplified model approach~\cite{LHCNewPhysicsWorkingGroup:2011mji,Gutschow:2012pw}, with a non-exhaustive discussion of how such final states are motivated by specific SUSY-breaking models. The typical search for SUSY, including many of those shown in Figure~\ref{fig:vanilla_susy}, targets pair-production of sparticles with potentially long cascades through lower-mass sparticles until the stable LSP is reached~\cite{Baer:1986au}. This results in multiple visible decay products and a characteristic large missing transverse momentum (\met) signature from the stable LSPs escaping undetected. The next-to-lightest supersymmetric particle (NLSP) in a specific model, which usually decays to the LSP and a SM particle, plays an important role in collider phenomenology. In the following subsections, we assume that R-parity is conserved. We define R-parity and discuss models with R-parity violation in Section~\ref{sec:rpv}.

\subsection{Semi-compressed and compressed regimes}
\label{sec:compressed}

Qualitatively, the shapes of the exclusion curves in Figure~\ref{fig:vanilla_susy} share a notable characteristic: the sensitivity in $m_{X}$ is maximum where $m_{\tilde{\chi}_1^0}$ is massless. For a heavy neutralino, the sensitivity to $m_{X}$ tends to nearly plateau at a fixed $m_{\tilde{\chi}_1^0}$ (for many $\tilde{q}$ and $\tilde{g}$ models) or to increase with $m_{X}$, staying away from the diagonal (for many slepton $\tilde{l}$ and EWKino $\tilde{\chi}$ models.) 

For $\tilde{q}$ and $\tilde{g}$ production, especially in all hadronic final states, this shape can be understood as a confluence of several factors. The signal is primarily identified by large total transverse energy in the event ($H_T$) and large \met. The SM multi-jet and $X$+jets (where $X$ = top, $W$, and $Z$) backgrounds are fairly continuous in both these variables, such that the sensitivity shape comes from changes in signal topology. The energy available to divide among decay products of a single sparticle, in its rest frame, is given by $\Delta m = m_{X} - m_{\tilde{\chi}_1^0}$; as $\Delta m$ decreases, both jet multiplicity and $H_T$ drop. For heavier $\tilde{\chi}_1^0$, \met\ also drops. While we may naively expect a heavier $\tilde{\chi}_1^0$ to carry away \emph{more}, it's important to note the difference here between mass and momentum. A heavy $\tilde{\chi}_1^0$ carries away energy in mass, but this is unmeasurable. All we can measure is the visible \emph{momentum} of the remaining SM decay products, which should match the \emph{momentum} of the invisible particles, in the transverse direction, and both will have lower momentum as the $\tilde{\chi}_1^0$ becomes heavier.

As long as $\Delta m$ is large, signal events, even those produced near rest, can easily produce \met\ and $H_T$ that are greater than strict selection requirements imposed to suppress the background. In this regime, the analysis sensitivity is, to first approximation, independent of $m_{\tilde{\chi}_1^0}$, which yields a near-vertical curve in cross-section sensitivity for a heavy $m_{X}$. As $\Delta m$ decreases and one enters the compressed or semi-compressed regime, both $H_T$ and \met\ from near-threshold $XX$ production drop below the most aggressive selection thresholds.

With no specific priors, the semi-compressed region is just as important an experimental target as the region with a massless \N{1}. Some acceptance is gained from Lorentz-boosted $XX$ production, where initial $p_T$ from the $X$ can be inherited by its decay products. In particular, the presence of initial-state radiation (ISR) is an important handle which, when present, boosts the recoiling sparticle pair system. This increases the $p_T$ and therefore the $A \times \varepsilon$ for the sparticle decay products. ISR-boosted production also increases the measured \met\ by increasing the visible (and imbalanced) momentum, potentially \textit{promoting} events into a high-scale signal region. Where the acceptance loss from requiring boosted $XX$ production hits the cross-section limit, the sensitivity turns over. The semi-compressed regime is a corner of model space in which \textbf{rapid optimization variation}, as a function of signal target, is required to balance the relationship between $\Delta m$, the production kinematics, and the increasing signal production cross-section as $m_{X}$ drops. To probe the semi-compressed regime as fully as possible, analyses must introduce many dedicated signal regions with varying selection requirements on \met, $H_T$, and related quantities~\cite{ATLAS:2020syg,CMS:2019zmd}.

\begin{marginnote}[]
\entry{ISR $p_T$ promotion}{Hadronic Initial State Radiation boosts the sparticle system and provides an additional jet in the event. The boost increases the measured $p_T$ of the sparticle decay products, and the additional jet increases both $H_T$ and \met.}
\end{marginnote}

For $\tilde{t}$, $\tilde{\chi}$, and $\tilde{l}$ production, accessing the semi-compressed and compressed regime faces the same signal kinematic challenges as for $\tilde{g}$ and generic $\tilde{q}$ production, with the additional challenge that the dominant background processes have massive electroweak bosons. Near-threshold production of $t\bar{t}$ and di-boson processes produce background events with kinematics (including \met\ and lepton $p_T$ for leptonic decays, or jet $p_T$ for kinematic decays) that mimic the signal kinematics for $\Delta m = m_{SM}$, where $m_{SM}$ is the massive SM object. Therefore, not only does the signal lose acceptance as one approaches the diagonal, but there is a regime in which the SM background is significantly enhanced due to threshold production. Boosting the $XX$ system boosts the background in a similar way, creating a particularly challenging corridor for these searches.

In addition to adding a sheer volume of signal regions, dedicated techniques have been developed to target the semi-compressed corners of simplified-model MSSM SUSY space. As discussed above, an $XX$ system boosted by an ISR jet helps promote the $p_T$ of the sparticles and improves background discrimination. However, the ISR jet adds an additional object to the event, increasing both the kinematic complexity of the reconstruction, as well as the potential to separate signal and background by the kinematic relationship of the visible jets, not just their number and magnitude. The \emph{Recursive Jigsaw Reconstruction} (RJR) technique was developed to analyze events in the presence of unmeasured (invisible) decay products and combinatoric unknowns associated with indistinguishable particles~\cite{PhysRevD.96.112007}, and it was extended during Run 2 of the LHC to specifically target ISR-boosted sparticle production by tagging the ISR jet and improving reconstruction of the boosted system despite unknown sparticle masses~\cite{ATLAS:2019lng, CMS:2025ttk}. Near the end of Run 2 and into Run 3, Machine Learning (ML) techniques are increasingly used to achieve multivariate separation between signal and huge SM backgrounds without explicit reconstruction (see \textit{e.g.} Ref.~\cite{CMS:2021eha}). 

As a signal becomes more compressed ($\Delta m \approx 1 -10$~GeV), the loss in signal acceptance originates not only from selection requirements at the analysis level, but also from object reconstruction and identification. Targeting low $p_T$ objects faces significant \textbf{detector limitations}, as the detectors are optimized to efficiently detect the multi-GeV $p_T$ particles from a typical $W$, $Z$, top, or $H$ decay.  Over the course of Run 2, the ATLAS and CMS collaborations have targeted moderately compressed regions for stop and sbottom production with dedicated soft b-tagging algorithms~\cite{ATLAS:2024ytk,CMS:2017mbm} and developed custom very low $p_T$ lepton reconstruction and identification techniques for compressed electroweakino and slepton searches, as discussed in Section~\ref{sec:higgsinos}. In the very compressed regime as $\Delta m \rightarrow 0$, the available phase space for decay becomes so restricted that the hard-scatter sparticle can become measurably long-lived, as discussed in Section~\ref{sec:sleptons}.

\begin{marginnote}[]
\entry{Soft di-$b$ + \met}{Two reconstructed low $p_T$ jets tagged as containing a $B$-hadron decay, plus significant \met. Reconstructing and identifying $b$-tagged jets with $p_T \lesssim 20$~GeV requires custom algorithms and dedicated calibrations.}
\end{marginnote}

\subsection{Electroweakinos} 

The superpartners of the EW gauge bosons (the bino $\tilde{B}$ and winos $\tilde{W}$) and of the Higgs (higgsinos) mix to form two charged mass eigenstates (charginos, $\tilde{\chi}_{1,2}^{\pm}$) and four neutral mass eigenstates (neutralinos, $\tilde{\chi}_{1,2,3,4}^0$) and are collectively called ``electroweakinos." In the MSSM, the soft mass parameters of the bino and winos are $M_1$ and $M_2$, respectively, while the higgsino mass parameter $\mu$ is unrelated to the SUSY-breaking scale~\cite{Bae:2019dgg}. If the bino, wino, and higgsino mass gaps are substantial, there is little mixing in the sector and the neutralino and chargino mass eigenstates are nearly pure bino, pure wino, and pure higgsino~\cite{primer}.

The lightest neutralino, \N{1}, is a prime dark matter candidate if stable, and in many models of natural supersymmetry the electroweakinos may be the only superpartners with masses accessible at the LHC~\cite{theoristSUSY_review,Baer:2012up}. Therefore, although their production rate at the LHC is smaller than for comparably-massive gluino and squarks because they can not be produced by strong interactions, electroweakinos are an important experimental target. Most signatures depend heavily on the mass ordering and the mass differences between the electroweakinos, and a comprehensive electroweakino search program is therefore challenging due to \textbf{luminosity limitations}, \textbf{rapid signature variation}, and \textbf{rapid optimization variation}. Notably, some signatures do not rely much on the electroweakino mass differences, including mono-jet searches and searches for electroweakinos that decay through an R-parity Violating coupling (see Section~\ref{sec:rpv}).

For a recent complete review of searches for electroweakinos, please see Ref.~\cite{Canepa:2020ntc}. In the following section, we focus on a well-motivated and particularly challenging electroweakino target: natural Higgsinos. 

\subsection{Natural Higgsinos}
\label{sec:higgsinos}

In measures of naturalness, the higgsino mass parameter $\mu$ enters at tree level, and therefore the higgsino masses are one of the most constraining naturalness parameters, generally required to be $\lesssim$ 300~GeV. In nearly all SUSY models motivated by naturalness, the higgsinos are the lightest superpartners, and often significantly lighter than the other electroweakinos~\cite{theoristSUSY_review, PhysRevD.91.055008, Baer:2011ec}. 

Before the discovery of the Higgs with a mass of 125~GeV, Anomaly-Mediated SUSY Breaking (AMSB) was a popular model of SUSY-breaking, with a specific prediction that the winos would be the lightest superpartners and an excellent dark matter candidate. The difficulty of generating a Higgs mass of 125~GeV within an AMSB model and other naturalness constraints within the original framework motivated a generalized natural model still mediated by anomalies, natural AMSB (nAMSB)~\cite{Baer:2018hwa}. In the nAMSB framework, higgsinos are the lightest sparticles, although the winos are still the lightest gauginos.

In such models, the higgsino triplet (\N{1}, \N{2}, \C{1}) may be the only superpartners with masses kinematically accessible at the LHC. If so, the experimental program is highly sensitive to their mass differences. Following Ref.~\cite{primer} and making the assumption that $M_1 = M_2 \gg |\mu|$, as well as the approximation that $m_Z \approx m_W$, the tree-level mass difference between the lightest chargino and neutralino, \deltaMpm[tree], can be approximated as:

\begin{equation}
\deltaMpm[tree] \approx \frac{m_Z^2( M_2 - 2|\mu|)}{M_2^2}
\end{equation}

\noindent when we only consider the leading terms and assume large $\tan{\beta}$, the ratio of Higgs VEVs. The first thing we note is that if $M_2$ (and $M_1$) are infinite, the higgsinos are mass degenerate at tree level. In this case, mass splittings of the order $\deltaMpm[loop] \approx 280-350$~MeV arise only from loop electroweak effects~\cite{IBE2013252,PhysRevLett.81.34}.  

If we take large but not infinite values of $M_2$ and $M_1$, we find that \deltaMpm[tree] scales like $\sfrac{1}{M_2}$. Therefore, the total mass difference, $\deltaMpm = m_{\C{1}} - m_{\N{1}} = \deltaMpm[tree] + \deltaMpm[loop]$, is fairly sensitive to $M_2$ and $M_1$ even if the winos and bino are an order of magnitude or two heavier than the higgsino. For example, for $M_2$ and $M_1$ fully decoupled, $\deltaMpm \simeq 300$~MeV for a higgsino triplet with mass of 200~GeV. For $M_2 = M_1 = 5$~TeV, \deltaMpm\ increases to over 1~GeV. This modest increase in \deltaMpm\ has important phenomenological implications, as discussed in the following. 

\begin{figure}
\includegraphics[width=5in]{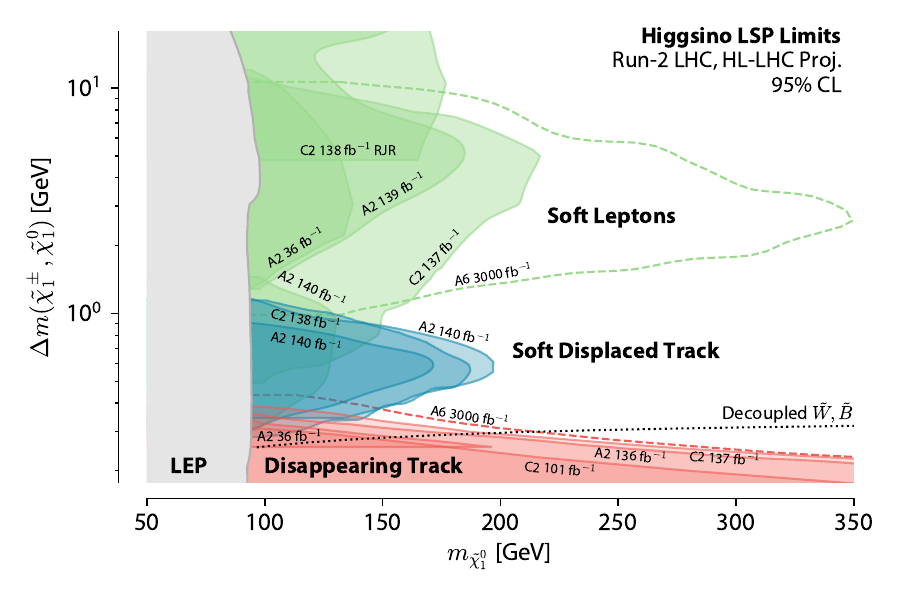}
\caption{The observed limits at 95\% CL for simplified-model Higgsino production, for $\deltaMpm = \Delta m(\C1,\N1)$ as a function of $m_{\N{1}}$. Results from LEP, ATLAS, and CMS are shown in solid colors for different signatures, while the dashed lines illustrate conservative projections for the HL-LHC dataset~\cite{ATLAS:2019lng, ATLAS:2021moa,ATLAS:2019lng,ATLAS:2022rme,CMS:2023mny,CMS-PAS-EXO-23-017, CMS-PAS-SUS-24-003, ATLAS:2018jjf, ATLAS:2024umc, ATLAS:2025lhc}. 
}
\label{fig:higgsino}
\end{figure}

The compelling naturalness motivations for light higgsinos and/or light winos have driven a comprehensive program at the LHC to cover their potential experimental signatures. Natural higgsinos are a textbook example of rapid signature variation, as small changes in \deltaMpm\ lead to big changes in the optimal search strategy. Moreover, each of the several different signatures are challenging in qualitatively different ways. The progress that first LEP and now the LHC have made in covering this SUSY space is a testament to decades of experimental and phenomenological effort. For the remainder of this section, we will review the signature parameter space as a function of \deltaMpm\ and discuss a few notes about future prospects. While we focus the discussion on higgsinos, the signatures are similar for the case in which the wino doublet is instead much lighter than the other electroweakinos. The current experimental sensitivity for Higgsino production is summarized in Figure~\ref{fig:higgsino}.

\subsubsection{Disappearing Tracks}

The minimum \deltaMpm\ consistent with loop radiative electroweak corrections is $\deltaMpm \approx 300$~MeV for Higgsinos and $\deltaMpm \approx 260$~MeV for winos~\cite{PhysRevLett.81.34,IBE2013252}, corresponding to a chargino lifetime of about 0.05~ns and 0.2~ns, respectively. At these \deltaMpm, the \C{1} would decay primarily to a neutralino and a pion with $p_T \approx 200-300$~MeV. This produces the canonical ``disappearing track" signature, in which the chargino traverses enough layers of the tracking detector to be reconstructed as a short, high-$p_T$ track before ``disappearing" to a neutralino and soft pion, neither of which are typically reconstructed. ATLAS and CMS have both designed custom tracking setups for reconstructing these short tracks, and both experiments have published searches for disappearing tracks that have steadily marched first through the wino parameter space and now, in Runs 2 and 3, have extended into Higgsino parameter space. As shown in Figure~\ref{fig:higgsino}, the current sensitivity reaches to about 200~GeV for a fully-decoupled higgsino~\cite{ATLAS:2022rme,CMS:2023mny}. 

\begin{marginnote}[]
\entry{Disappearing Track}{A short track from a charged long-lived particle that decays into unreconstructed particles after traversing several layers of the tracker.}
\end{marginnote}

Searches for disappearing tracks are fundamentally limited by the short chargino lifetime, the fact that their decay position samples from an exponential for a given boost, and by detector limitations: collider detectors are not optimized for reconstructing charged LLPs in this lifetime range. The higgsino-motivated disappearing track has a severe acceptance penalty when requiring that the chargino traverse enough detector layers to be reconstructed. Current results from ATLAS and CMS require the \C{1} to traverse at least 4 detector layers. With the 4th layer of the trackers at 122.5~mm and 160~mm for ATLAS and CMS, respectively, the acceptance as a function of decay position, $A(d) = e^{-\frac{d}{\beta\gamma c \tau}}$, is roughly $0.03\%$ and $0.002\%$ for $\beta\gamma =1$ and a lifetime $\tau = 0.05$~ns. 

To increase analysis sensitivity, experiments must identify novel ways to reconstruct shorter tracks, while maintaining background rejection against combinatoric tracks from unassociated hits, whose number explode as the number of hits drops. ATLAS has tackled this challenge by pursuing 3-layer track reconstruction~\cite{ATL-PHYS-PUB-2019-011}, for which the acceptance increases from $0.03\%$ to $0.27\%$. To keep backgrounds levels in-check, the experiment reconstructs the displaced, low-$p_T$ track from the pion which emerges at the end of the chargino track~\cite{ATL-PHYS-PUB-2019-011}. An additional challenge is that the momentum reconstruction becomes unstable with 3 or fewer hits; one solution is to use the primary vertex position as an additional track constraint, trading any knowledge of the impact parameters for improved momentum resolution, in the case that the track actually did emerge from the primary vertex~\cite{kaji}.

Searches for disappearing tracks are especially challenging for the LHC program because they are one signature for which the new tracking detectors designed for the HL-LHC pose additional challenges, rather than clear benefits. The optimal disappearing track detector is in tension with conventional tracking needs, for which widely-spaced silicon layers improve resolution on the helical tracking parameters. In particular, relative to the current ATLAS inner detector, the four innermost layers of the future ATLAS Inner Tracker (ITk) will be at larger radii, with the 4th at 228~mm~\cite{ATLAS:2019mta}, resulting in an acceptance for 4-layer higgsino tracks that is 3 orders of magnitude smaller than the current detector.  A conservative extrapolation of the Run-2 analysis clearly shows the effect of this acceptance loss: the HL-LHC projection shows that the $5\sigma$ discovery potential for a full $3000$~fb$^{-1}$ of data with ITk reaches roughly the same sensitivity as the region excluded by the Run-2 $140$~fb$^{-1}$ ATLAS search~\cite{ATLAS:2018jjf}. To extend the discovery potential for natural higgsinos at the HL-LHC, the experiments must continue to push to shorter tracks. It remains to be seen whether or not, for example, a 2-layer track, combined with a vertex-constraint and the reconstruction of a low-$p_T$ pion, has enough separation power against enormous combinatoric backgrounds to improve sensitivity~\cite{Fukuda:2017jmk}. Other handles may be possible, including perhaps requiring a Final-State Radiation (FSR) photon~\cite{Ismail:2016zby} or requiring above-MIP ionization for the hits from the chargino. CMS has already employed the latter strategy~\cite{CMS:2023mny}, although the gain is sensitivity is not clear for the current higgsino mass reach. 

\subsubsection{Soft di-lepton and \met\ signature} At $\deltaMpm \gtrsim 350$~MeV, the \C{1} lifetime becomes too short to efficiently reconstruct in the tracker, and we must instead search for the decay products of the \C{1} (for both the higgsino and wino case) and/or the \N{2} (for the higgsino case.) At $\deltaMpm \gtrsim 1$~GeV, current sensitivity is dominated by searches for events with leptons and \met. The leptons originate from the decays of $W$, $Z$, or $H$ bosons emitted in the decay of an electroweakino to a lighter electroweakino, and the kinematics of the leptons are highly sensitive to the mass parameters of the electroweakinos. ATLAS and CMS have targeted this regime by taking advantage of the kinematic end-point in the di-lepton invariant mass spectrum, $m_{ll}$, that arises from the decay of the off-shell $Z$ boson in the transition $\N{2}\rightarrow Z^* \N{1}$. At values of $\deltaMo = m_{\N{2}} - m_{\N{1}} \gtrsim 25$~GeV, requiring a third lepton from a $W^*$ decay in the $\C{1}\rightarrow W^* \N{1}$ transition brings additional sensitivity~\cite{ATLAS:2021moa}, but at smaller mass splittings, this lepton becomes too soft to efficiently reconstruct, especially if $m_{\N{1}} < m_{\C{1}} < m_{\N{2}}$, as is often assumed.

For $1 \lesssim \deltaMo \lesssim 25$~GeV, the dominant sensitivity to higgsino production arises from the soft di-lepton and \met\ signature~\cite{Baer:2011ec,Han:2014kaa,ATLAS:2019lng,CMS-PAS-EXO-23-017}. Note that both experiments assume that the $BR(\N{2}\rightarrow Z^* \N{1})=100\%$ and that $m_{\C{1}} = (m_{\N{1}}+m_{\N{2}})/2$, which neglects the possibility of $\N{2}\rightarrow W^* \C{1}$. While these analyses target the $\N{2}\rightarrow Z^* \N{1}$ decay in \N{2}\N{1} and \N{2}\C{1} production, additional sensitivity arises from \C{1}\C{1} production. It is interesting to note that both ATLAS and CMS currently observe small excesses at $\deltaMo \approx 5-30$~GeV~\cite{ATLAS:2019lng,CMS-PAS-EXO-23-017, Agin:2024yfs}.

\begin{marginnote}[]
\entry{Soft di-lepton + \met}{A pair of leptons with low transverse momentum, which is below the thresholds for efficient reconstruction and identification with standard algorithms. \met serves as both a trigger and handle for background rejection.}
\end{marginnote}

The soft di-lepton plus \met corner of signature-space faces challenges from the start: the acceptance from triggering on \met, which originates from ISR, and the di-lepton branching fractions combine to be less than $1\%$~\cite{ATLAS:2019lng}. Because the lepton kinematics vary strongly as a function of \deltaMo, both experiments have developed highly sophisticated event selections designed to optimize for kinematic changes as a function of $m_{ll}$. Experimentally, the soft lepton $p_T$ distribution poses an enormous challenge for efficiently reconstructing and identifying the leptons, as well as for rejecting enormous backgrounds without cutting too aggressively into the already-compromised acceptance. Both ATLAS and CMS have deployed custom lepton reconstruction and/or identification algorithms in pursuit of these very soft leptons: ATLAS developed a muon identification algorithm which significantly improved the efficient ID of muons down to 3~GeV specifically for this signature~\cite{ATLAS:2019lng}, while CMS used a low-$p_T$ electron reconstruction algorithm originally developed for tests of lepton flavor universality in $B^{±}$ decays, which allows electrons to be reconstructed down to $p_T = 1$~GeV~\cite{CMS:2024syx}. 

Algorithmic improvements can only be taken so far, however: muons with $p_T \lesssim 3$~GeV are unlikely to make it to the muon systems for either detector, and electrons with $E_T \lesssim 1$~GeV will rarely make it to the electromagnetic calorimeters. To address this, both experiments have developed signal regions which target one standard lepton and one track which is not matched to a fully reconstructed lepton. CMS trains a Boosted Decision Tree (BDT) to select the track most consistent with coming from a \N{2} decay while ATLAS employs a Deep Neural Network (DNN) model to identify low-$p_T$ electrons  with $0.5 < p_T < 5$~GeV and muons with $1 < p_T < 1$~GeV~\cite{CMS-PAS-SUS-24-003, ATLAS:2025lhc}.

\subsubsection{Soft displaced tracks} For more than 10 years after the start of data-taking, the LHC had nothing to say in the region $0.3 \lesssim \deltaMpm \lesssim 1$~GeV, and the best sensitivity belonged to the LEP legacy~\cite{LEP_charginos}, combining all four LEP experiments and integrating over several signatures as a function of decreasing \deltaMpm, including prompt leptons, soft events with photon ISR, tracks with displaced impact parameters, kinked tracks, and searches for heavy, stable, charged particles using a time-of-flight or ionization signature. Indeed, in this particularly challenging \deltaMpm\ region, the LEP sensitivity arose from searches for tracks with displaced impact parameters, and it was relatively recently recognized that the LHC would have unique sensitivity in a similar signature: high \met\ plus a low $p_T$, displaced track~\cite{Fukuda:2019kbp}. While this regime is a small fraction of the higgsino \deltaMpm\ phase space if sampled democratically, it is phenomenologically important as it emerges in scenarios with natural higgsinos but not fully decoupled winos and binos ($5 \lesssim M_1 = M_2 \lesssim 100$~TeV). It is also experimentally important, as direct dark matter experiments have limited sensitivity in this regime due to vanishing DM-nuclei couplings~\cite{Fukuda:2019kbp}.

\begin{marginnote}[]
\entry{Soft displaced track + \met}{A low-momentum track with a measurably displaced transverse or longitudinal impact parameter with respect to the primary vertex, from a charged pion or lepton which emerges in the decay of a slightly long-lived heavy parent. An additional heavy and neutral decay product escapes the detector without interacting.}
\end{marginnote}

In the last two years, both ATLAS and CMS have released searches targeting this isolated, low $p_T$, mildly displaced track plus \met\ signature~\cite{ATLAS:2024umc, CMS-PAS-SUS-24-012, ATLAS:2025lhc}. For $0.3 \lesssim \deltaMpm \lesssim 1$~GeV, the chargino acquires a lifetime of $O(0.1-1)$~mm. This \C{1} lifetime is prohibitively inefficient to reconstruct as a track, but it can be targeted by reconstructing the track from the pion or lepton from the $\C{1}\rightarrow\N{1}\pi^{\pm}$ decay (and to a lesser extent, but still included, are displaced tracks from $\N{2}\rightarrow\N{1}l\bar{l}$ decays). This track emerges at a displaced distance and can be targeted by identifying the significance of its impact parameters with respect to the primary vertex of the event. As no lepton is required, the decay products can be much more efficiently reconstructed than in the soft di-lepton signature: the tracking of both experiments is fully (90\%) efficient for pions with $p_T > 1$~GeV, and nearly fully efficient (80\%) for pions with $p_T > 500$~MeV. Rather, there emerges an experimental challenge from targeting low $p_T$ tracks not because they are difficult to reconstruct but because there are too many of them: by default ATLAS only saves information in the standard analysis format for tracks with $p_T > 1$~GeV to significantly reduce the size of the stored and processed data. Targeting events with $0.1 < p_T < 1$~GeV requires custom data processing and data formats, which can be cumbersome. This is an example of \textbf{computing constraints} in targeting non-conventional signatures. In addition to data processing limitations, the enormous multiplicity of low-$p_T$ pions, many of which are displaced due to hadron decays, requires additional handles to reject background. Both ATLAS and CMS again rely on a large \met\ selection to reject backgrounds. Additionally, there is an almost-irreducible background from $W\rightarrow\tau\nu$ events, due to the measurable $\tau$ lifetime, although placing an upper limit on the track $p_T$ reduces it. Both ATLAS and CMS use targeted Neural Networks to discriminate using track displacement parameters as well as event-level variables~\cite{CMS-PAS-SUS-24-012, ATLAS:2025lhc}.

\subsubsection{Additional signatures with \met} Across the whole natural higgsino landscape, current searches use a \met\ trigger. At trigger-level, the \met\ originates from the combination of an ISR jet which is reconstructed and superpartners which are not. Offline, most higgsino searches require a tighter selection on \met\ to reject backgrounds. It is logical to wonder if the purely \met-based mono-jet signature would add sensitivity at some \deltaMpm. Thus far, due to the relatively large $W$ and $Z$ backgrounds which remain even after tight mono-jet requirements, and due to the theoretical and experimental uncertainties that accompany these backgrounds, it is not yet clear if the mono-jet signature (limited by \textbf{systematic uncertainties}) alone will be sufficiently powerful to add sensitivity to higgsino production, even at the HL-LHC~\cite{Han:2013usa, Baer:2014cua}. 

\begin{marginnote}[]
\entry{Mono-jet}{A single jet with large transverse momentum. In searches for supersymmetry, the jet usually originates from ISR gluon or quark emission, and unreconstructed superpartners carry away momentum which appears as \met.}
\end{marginnote}

To continue to drive new sensitivity at the LHC and HL-LHC, the experiments must continue to push signatures which combine \met\ plus additional, higgsino-specific handles. The recent sensitivity gains shown by both ATLAS and CMS with soft, displaced tracks illuminates the power of new signatures, especially those which have a large intrinsic signal acceptance. Looking forward, several novel ideas have been proposed to continue to explore this important parameter space, including taking advantage of the fact that charginos are electrically charged to target \met\ plus Final State Radiation (FSR) photons~\cite{Ismail:2016zby}. Once large \met\ is required, many of the main backgrounds, such as $Z\rightarrow\nu\nu$, may emit ISR photons but won't emit FSR photons aligned with the \met. It remains to be seen if the large production of relatively soft photons can be reduced to a manageable level, but the photon plus \met\ signature has the advantage of being relatively independent of both \deltaMpm\ and the exact decay assumptions of the higgsino states. Moreover, targeting a photon plus \met\ final state where the photon emerges from the electroweakino decay would allow for more comprehensive coverage of model parameters~\cite{Baum:2023inl}.

\begin{marginnote}[]
\entry{photon + \met}{A single, isolated photon plus significant \met. The photon could be produced by initial or final-state radiation, or from electroweakino decay, targeting different superpartner production and decay, with different kinematics and different backgrounds.}
\end{marginnote}

Additionally, as discussed in Section~\ref{sec:sleptons}, photon-induced electroweakino pair-production is an attractive target, as the protons in such events could in principle be measured by forward detectors, tagging the total missing energy~\cite{Khoze:2017igg}, although it is not clear how much sensitivity this channel would add on top of the continually-improving searches for soft leptons and displaced tracks~\cite{Zhou:2024fjf}. Similarly, we propose that photon-induced pair-production of \C{1} states in ultra-peripheral heavy-ion collisions is a promising avenue to pursue, especially for relatively light ion species. Despite the low integrated luminosity of the heavy-ion datasets, the production cross-section is enhanced by a factor of the beam particle charge $Z^4=\mathcal{O}(10^7)$ for ions due to the intense photon field surrounding each of two projectiles, and the clean environment of these events can allow suppression of the hadronic backgrounds that otherwise require acceptance-poor selections, like high \met, in $pp$ collisions. 

\subsection{Sleptons} 
\label{sec:sleptons}

For simplified-model MSSM production, sleptons are the rare birds. Like charginos and neutralinos, sleptons can only be produced via electroweak interactions. However, unlike the fermionic electroweakinos, the scalar sleptons must be produced in a state of total angular momentum = 1 to conserve spin, which significantly suppresses slepton production, especially near threshold. The production cross-section of sleptons is roughly an order of magnitude smaller than higgsinos for the same sparticle mass. An additional complication is that the super-partners of the left-handed and right-hand leptons, $\tilde{l}_L$ and $\tilde{l}_R$, respectively, could have different masses, and their mixing is unknown, although as the mixing is proportional to the corresponding fermion mass, it is expected to be small~\cite{Baer:2006rs}. In this work, we ignore the subtleties of potential $\tilde{l}_R$ and $\tilde{l}_L$ mixing, as it would affect the production cross-section but not the resulting signatures; we will focus our discussion on the case in which there is no mixing. While $\tilde{l}_L$ are SU(2)$_L$ doublets, $\tilde{l}_R$ are SU(2)$_R$ singlets, so they only couple to the $Z$ boson via hypercharge. This reduces the s-channel $Z/\gamma^*$ production of $ \tilde{l}_R$ to roughly $30-40\%$ of the $\tilde{l}_L$ s-channel production~\cite{Bozzi:2007qr, Fuks:2013lya, PhysRevLett.83.3780}. $\tilde{l}_R$ are the rarest of the rare birds.

Despite their elusiveness, relatively light sleptons ($\slepton[] \lesssim 1$~TeV) are well-motivated. The bino is a classic LSP candidate, due to dark matter considerations, implications from grand-unified theory investigations, and factors that arise in specific SUSY-breaking models~\cite{HABER198575, Jungman:1995df, Baer:2006rs}. However, if the LSP is a bino-like neutralino, it is easy to produce too many \N{1} to match the dark-matter relic density. Slepton-neutralino co-annihilation in the early universe, which requires a small mass difference between the slepton and the neutralino, could have reduced the dark matter relic density to the observed value~\cite{Aboubrahim:2017aen,Ellis:1998kh,Ellis:1999mm}, thus motivating that the mass of the \N{1} and the lightest \slepton\ are both $\lesssim 1$~TeV. 

Simplified-model slepton production in which pair-produced \slepton\ each decays through the channel $\slepton\rightarrow \ell \N{1}$ has long been a collider target and is the focus of our discussion. Unless we specify a SUSY-breaking mechanism, however, the relationship between the three slepton flavors, \selectron, \smuon, and \stau, is unknown, as is the relationship between the mass of the \slepton[R] and \slepton[L] for each flavor, although RG running effects favor $m_{\slepton[R]} < m_{\slepton[L]}$. The experimental sensitivity to $m_{\slepton}$ depends heavily on the assumptions of the mass relationship between the sleptons. However, the experimental sensitivity to $\sigma_{\slepton}$ is independent of these assumptions, and instead depends heavily on $\Delta m_{\slepton} = m_{\slepton} - m_{\N{1}}$. Often, experiments will perform multiple interpretations, including, for example, degenerate cases such as $m_{\selectron[R]} = m_{\selectron[L]}$. 

The LEP experiments excluded \selectron[R], \smuon[R], and \stau[R] below masses of 100, 95, and 65~GeV for $\Delta m_{\slepton} \gtrsim 2$~GeV~\cite{LEP_sleptons}, and ALEPH uniquely set a lower limit $m_{\selectron[R]} \gtrsim 75$~GeV, independent of $\Delta m_{\slepton}$~\cite{ALEPH:2002nwp}. While LEP made no explicit interpretation in terms of \slepton[L], this is because they assumed that $m_{\slepton[L]} > m_{\slepton[R]}$; the searches were signature-agnostic to \slepton[R] or \slepton[L], and can be treated as well as bounds on $m_{\slepton[L]}$. There also exists a hard constraint that $m_{\slepton} > 40$~GeV from the LEP measurements of the $Z$-width, for each individual slepton type~\cite{ParticleDataGroup:2024cfk}.

If $\Delta m_{\slepton} \gtrsim m_{W}$, searches target two opposite signed, same flavor leptons with moderate \met. In this case, the leptons are directly trigger-able, and reasonable $A\times \varepsilon$ of $O(25\%)$ or more can be achieved at the LHC for \selectron\ and \smuon\ production~\cite{ATLAS:2019lff,ATLAS:2025evx}, and order $O(5\%)$ for hadronically-decaying \stau~\cite{ATLAS:2024fub}. For light \N{1}, the current LHC sensitivity extends to about 600~GeV for mass-degenerate \selectron[R] and \selectron[L], as well as for mass-degenerate \smuon[R] and \smuon[L]~\cite{ATLAS:2019lff,CMS-PAS-SUS-23-002}. For $m_{\stau[R]} = m_{\stau[L]}$, masses up to 500~GeV are excluded for light \N{1}~\cite{ATLAS:2024fub}. In the case of solo $\stau[R]\stau[R]$ production, masses between $100-350$~GeV are excluded.

As $\Delta m_{\slepton}$ decreases, lepton $p_T$  and \met\ drop, decreasing both signal acceptance as well as rejection power against  backgrounds for the soft di-lepton plus \met\ signature. Moreover, around $\Delta m_{\slepton} = m_W$, the background from $W^+W^-$ production increases rapidly, producing an especially challenging compressed corridor. This regime is a great target for advanced ML techniques, which are powerful at working in the multi-dimensional kinematic space to separate signal from background with as little impact to signal acceptance as possible~\cite{Dutta:2023jbz}. 

In the last year, ATLAS  and CMS have made significant sensitivity gains into this compressed corridor, utilizing a \met\ trigger and a number of BDT algorithms, each specifically targeting a unique $\Delta m_{\slepton}$~\cite{ATLAS:2025evx,CMS:2025ttk}. The ATLAS collaboration also published results for a cut-and-count analysis targeting the same phase space. To reach reasonable signal significance, the cut-and-count analysis reaches a $A\times\varepsilon$ of $2\times10^{-4}$ for \smuon\ with $\Delta m_{\slepton} = 5$~GeV. The BDT channel for the same target increases the $A\times\varepsilon$ by a factor of 20. For such a rare signal, with a very large SM background, the one-dimensional cut-and-count approach can not reduce the background to an acceptable level without suffering too significant acceptance losses. For the 4-fold degeneracy assumption of $m_{\selectron[R]} = m_{\selectron[L]} = m_{\smuon[R]} = m_{\smuon[L]}$, the BDT-approach excludes sleptons up to masses of nearly 200~GeV for $\Delta m_{\slepton} \gtrsim 3$~GeV. Still, ML is not magic. For sole $\selectron[R]\selectron[R]$ or $\smuon[R]\smuon[R]$ production, the BDT approach only excludes a small sliver of masses beyond LEP, and only for $3 \lesssim \Delta m_{\slepton} \lesssim 5$~GeV. Note that both ATLAS and CMS observe excesses in Run-2 searches targeting \smuon\ production with $\Delta m_{\smuon} \lesssim 10$~GeV~\cite{CMS:2025ttk,ATLAS:2025evx}.

\subsubsection{$\tau$-sleptons} 
\begin{figure}
\includegraphics[width=6.0in]{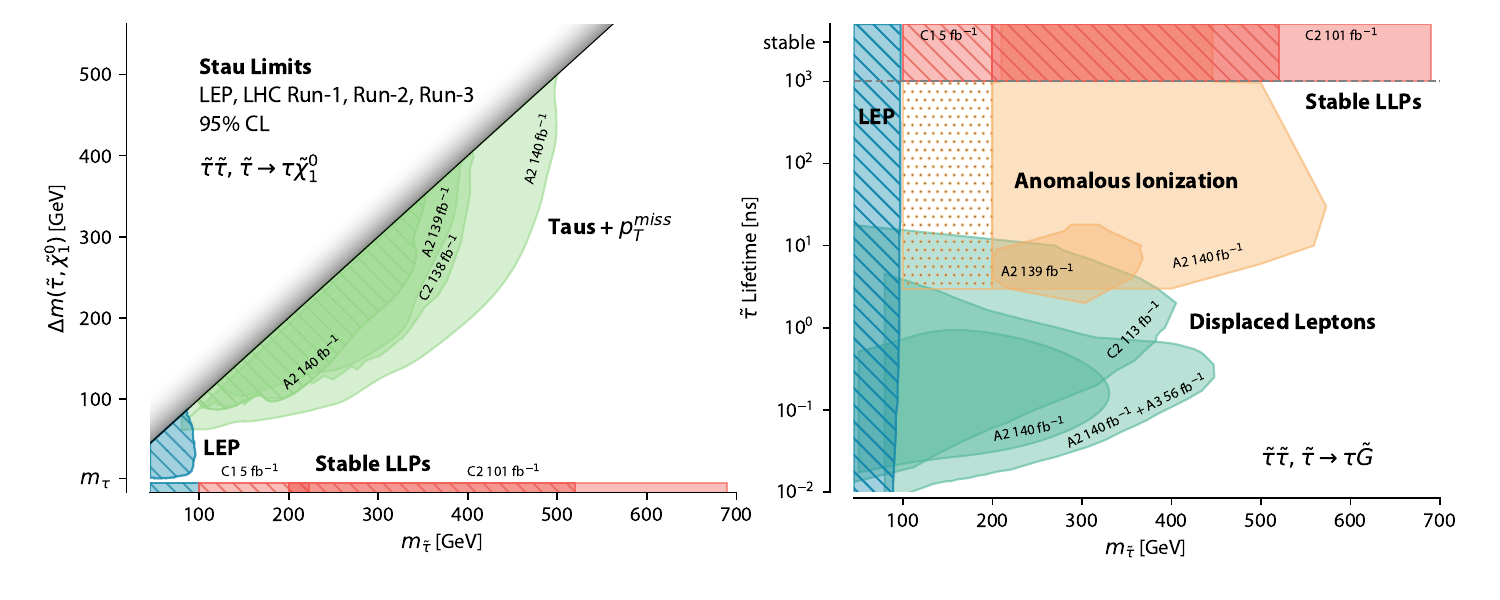}
\caption{The observed limits at 95\% CL for simplified-model $\stau\stau$ production, as a function of $m_{\stau}$. On the left, the limits for the MSSM $\stau\rightarrow\tau\N1$ process are shown with varying $\Delta m_{\stau} = m_{\stau} - m_{\N1}$. On the right, limits for GGM $\stau\rightarrow\tau\tilde{G}$ decays with a massless $\tilde{G}$ are shown as a function of $\stau$ lifetime, and are discussed in Section~\ref{sec:gravitinos}. Hashed lines indicate exclusion to solo $\stau[R]\stau[R]$ production, while filled areas exclude the degenerate case where $m_{\stau[L]} = m_{\stau[R]}$ and both are produced\protect\footnotemark~\cite{LEP_sleptons,LEP_sleptons_gmsb,ATLAS:2019gti,CMS:2012wcg,ATLAS:2024fub,CMS:2024nhn,CMS:2022syk,ATLAS:2020wjh,CMS:2021kdm,ATLAS:2022pib,ATLAS:2025fdm,ATLAS:2024vnc}.
}
\label{fig:staus}
\end{figure}

In many SUSY models, the \stau\ is predicted to be lighter than the \selectron\ and \smuon. Even if the soft mass parameters for all 3 sleptons are comparable, the larger Yukawa coupling for the $\tau$ creates stronger \stau[L] and \stau[R] mixing and tends to drive the renormalization-group running of the \stau\ to lighter masses than its counterparts~\cite{Barger:1993gh, Baer:2006rs, Ellis:1999mm}. Due to the inherent $A\times\varepsilon$ losses in reconstruction and identification of taus, searches for $\stau$ via the di-tau + \met\ signature face additional challenges relative to searches for $\selectron$ and $\smuon$. Additionally, due to the higher $p_T$ threshold requirements for $\tau$ triggers, searches for direct $\stau$ production face strong \textbf{trigger limitations}. As shown in Figure~\ref{fig:staus}, the coverage of MSSM $\stau$ parameter space is growing, but there is still ample room for discovery. Indeed, the LHC has no reported sensitivity to direct \stau\ production in the range $m_{\tau} \lesssim \Delta m_{\stau} \lesssim 60$~GeV. Co-annihilation arguments suggest this is an important regime and motivates dedicated searches. 

\footnotetext{The orange dotted region was not explicitly excluded by a dedicated search, although it is effectively excluded by a validation region of Ref.~\cite{ATLAS:2025fdm}. For the stable LLP interpretations as a function of $\Delta m_{\stau}$, we assume the $\stau$ is detector-stable for $\Delta m_{\stau}\lesssim m_{\tau}$.}

\begin{marginnote}[]
\entry{Di-$\tau$ + \met}{Two hadronically-decaying $\tau$s plus significant \met. Requiring the hadronic decay mode increases the acceptance of the $\tau$, but it is challenging to efficiently trigger, reconstruct, and identify hadronic $\tau$-decays, and the background is large.}
\end{marginnote}

As the $\tau$ decay imposes an additional acceptance loss, it was proposed that a signature with only one reconstructed and identified $\tau$, plus \met\ and a single jet from ISR might extend sensitivity in the compressed \stau\ corridor~\cite{Florez:2016lwi}. CMS targeted this signature and obtained sensitivity for compressed $\stau$s produced via electroweakino decay~\cite{CMS:2019zmn}, but no sensitivity (yet) to direct $\stau$ production. For the moment, di-tau + \met\ searches dominate. The experiments must continue to push into the small $\Delta m_{\stau}$ regime, continuing to innovate at the trigger, reconstruction, and analysis level, potentially relying more on ISR and \met\ at the trigger and analysis-level, as the searches for compressed \smuon\ and \selectron\ have done. The recent gains in sensitivity to natural higgsinos emerging from an intense program targeting multiple signatures as $\Delta m$ varies provide a promising road-map for what can be achieved.

\begin{marginnote}[]
\entry{Soft $\tau$ + jet + \met}{A single, isolated, hadronically-decaying $\tau$ with low $p_T$ plus significant \met\ and a jet from ISR production.}
\end{marginnote}

\subsubsection{Compressed Sleptons} For $\Delta m_{\slepton}<2$~GeV, the story changes. Unlike the higgsino case, there is no apriori theory motivation for a very small mass splitting between the sleptons and \N{1}, but co-annihilation arguments and the  dark matter relic density suggest there may be observational motivations instead. Also unlike the higgsino case, there is no lower bound on $\Delta m_{\slepton}$. As $\Delta m_{\slepton}$ drops, the sleptons acquire a macroscopic lifetime, and this is particularly dramatic for the \stau\ which decays through the noticeably heavy $\tau$~\cite{Citron:2012fg}.\footnote{At $\Delta m_{\stau} = 2$~GeV and above, the lifetime of the stau, $\tau_{\stau}$, is essentially prompt at $10^{-20}$~sec or shorter. Just a hair below $\Delta m_{\stau} = m_{\tau}$, the lifetime increases dramatically to $\tau_{\stau} = 1$~ns~\cite{Citron:2012fg}. By $\Delta m_{\stau} = 1.4$~GeV, $\tau_{\stau} = 100$~ns, which is close to detector-stable at the LHC.} Although the experiments have not yet performed interpretations of searches for charged LLPs in terms of co-annihilation sleptons, they should have existing exclusion up to around $m = 560$~GeV for $\tau_{\slepton} \gtrsim 10$~ns from searches with anomalous ionization~\cite{ATLAS:2025fdm} and to $m=700$~GeV for $\tau_{\slepton} \gtrsim 1000$~ns using anomalous ionization and/or time-of-flight measurements~\cite{CMS:2024nhn} (see Section~\ref{sec:gravitinos}), assuming mass degeneracy between the \slepton[R] and \slepton[L] partners. We illustrate this sensitivity in Figure~\ref{fig:staus}. This sensitivity is largely independent of slepton flavor. We note that 1~ns is the sweet-spot for the disappearing track searches, and therefore, the LHC should already have sensitivity beyond LEP for the full range $\Delta m_{\stau} < m_{\tau}$. For \selectron\ and \smuon\ decay, the lifetime advantage only kicks in at much smaller $\Delta m_{\slepton}$, and therefore there remain gaps in sensitivity in this $\Delta m_{\slepton}$ sliver for \selectron\ and \smuon\ production.

In between the LLP-regime and regime that approaches $\Delta m_{\slepton} = m_W$, the searches optimized for natural higgsinos, targeting the low-$p_T$ di-leptons plus \met\ signature described in Section~\ref{sec:higgsinos}, add some sensitivity for \smuon\ and \selectron\ production with $2 \lesssim \Delta m_{\slepton} \lesssim 20$~GeV. However, due to the smaller production rate of \slepton\slepton\ relative to higgsinos, the signal regions targeting sleptons cannot cut as hard on \met\ without cutting out all the potential signal events. In this case, a different cross-section affects the search strategy, not only the interpretation. 

\subsubsection{Sensitivity and $A\times \varepsilon$} 
\begin{figure}
\centering
    \includegraphics[width=5.0in]{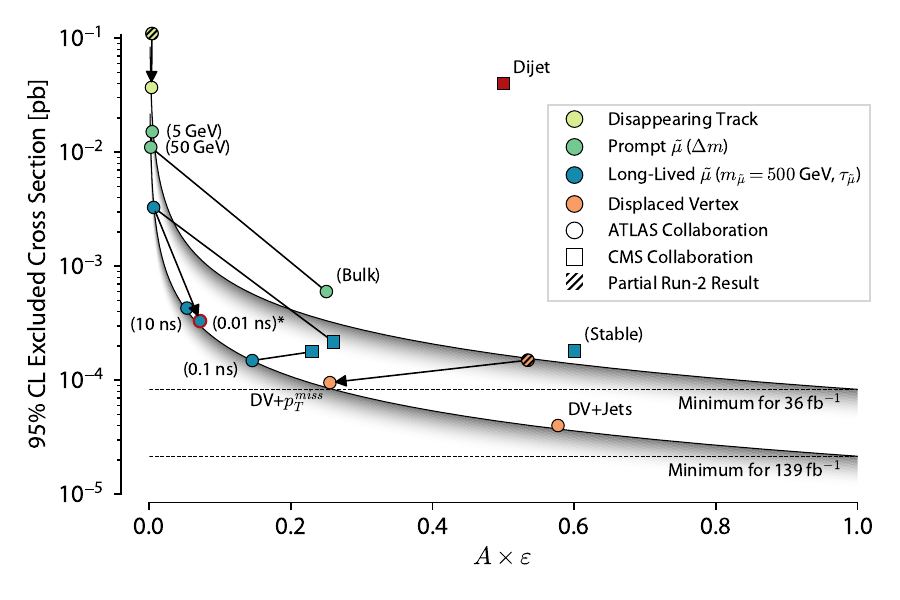}
\caption[Caption]{
Excluded cross-section results are shown from a collection of Run-2 analyses as a function of  the $A\times\varepsilon$ of the search. Lines represent similar signatures or model parameters, and arrows denote the evolution of an analysis. The shaded region shows the minimum possible cross-section exclusion for a given integrated luminosity and $A\times\varepsilon$.\protect\footnotemark~\cite{CMS:2019gwf,ATLAS:2023oti,ATLAS:2017tny,dvmetthesis,CMS:2024nhn,CMS:2021kdm,ATLAS:2020wjh,ATLAS:2023ios,ATLAS:2019lff,ATLAS:2025evx,ATLAS:2017oal,ATLAS:2022rme}
}
\label{fig:axe_plot}
\end{figure}

To illustrate the importance of $A\times \varepsilon$ in search strategy, and to highlight how the nature of a signature drives different working points, we find it helpful to compare the sensitivity of different searches as a function of their $A\times \varepsilon$ (see Figure~\ref{fig:axe_plot}). To compare searches for sparticles with different production cross-sections, we plot the sensitivity in terms of the cross-section (rather than mass) excluded at the 95\% CL.

Let's walk through a few examples in Figure~\ref{fig:axe_plot} to understand how different signatures manifest in this space. First, a search for a moderately heavy $Z'\rightarrow$di-jet resonance is relatively efficient to select (two well-reconstructed back-to-back jets, with an invariant mass within some small window), but this signature is limited by a large, irreducible di-jet background. Therefore, a di-jet search~\cite{CMS:2019gwf} can have large ($>50\%$) $A\times \varepsilon$~\cite{CMS:2019gwf}, but its sensitivity is many orders of magnitude worse than a background-free search at the same $A\times\varepsilon$, as illustrated by the solid curves in the plot which show the best (minimum) excluded cross-section for a background-free search with two different dataset luminosities, corresponding to roughly the partial Run-2 and full Run-2 datasets for the LHC experiments. Conversely, the ATLAS search for displaced vertices plus jets, targeting the decay of a long-lived, heavy, $R$-hadron (see Section~\ref{sec:rpv}), has a huge sensitivity advantage. Although the signature is unusual to reconstruct, and challenging for that reason, once one has gone through the work of building custom reconstruction and data processing algorithms, the signal would stick out like a lighthouse on a clear, dark night: the signal has very high (nearly 60\%) $A\times\varepsilon$ and the background is nearly zero. Of the whole LHC search program, this search reaches the best cross-section sensitivity (40 ab!) for its target model, as far as we are aware. To put this in context, this search probes processes rarer than SM $HH$ production (with cross-section $\approx$ 30~fb) and even SM $HHH$ production (roughly 100~ab)~\cite{Abouabid:2024gms}.

\footnotetext{The point with a red outline represents an analysis where the reported cross section exclusion was inconsistent with the reported $A\times\varepsilon$, but was re-calculable from the auxiliary material. This representation of a signal region's result serves as a useful cross-check.}

Now we come to the searches for sleptons in the 2D plane of Figure~\ref{fig:axe_plot}. For simplicity, we only include \smuon\ searches. For smuons with $\Delta m_{\slepton} > m_W$, we show a representative point from the ``bulk" region with $\Delta m_{\slepton} \approx 200$~GeV, where the search shows a somewhat standard SUSY search behavior: reasonable $A\times\varepsilon \approx 25\%$ with significant remaining background put the sensitivity about an order of magnitude above that of a background-free search at this working point. The cross-section sensitivity of this bulk smuon point, at slightly below 1~fb, is fairly typical of the current sensitivity of many LHC SUSY searches. Conversely, the compressed smuon corridor, due to the enormous $WW$ background, is forced to work in a much reduced $A\times\varepsilon $ regime. The very-compressed smuon corridor has additional intrinsic acceptance challenges for lower-$p_T$ muons. Interestingly, the BDT employed in the highlighted search optimizes for similar $\frac{S}{B}$ for the different $\Delta m_{\slepton}$ windows, but at slightly different $A\times\varepsilon$ working points~\cite{ATLAS:2025evx}. Once the smuons acquire a detector-stable lifetime of $\tau_{\slepton} \gtrsim 100$~ns, the signature completely changes, and the analysis jumps to a very high $A\times\varepsilon$ of above 60\%, with a corresponding increase in sensitivity. The complexity of a signature is not always correlated with the sensitivity. Further discussion on the $A\times\varepsilon$ behavior for searches for \smuon\ with intermediate lifetimes is continued in Section~\ref{sec:gravitinos}.

Conversely, the disappearing track signature as a search for higgsinos is limited by a brutal $A\times\varepsilon$ penalty out-of-the-box, imposed by the distance requirement for a short lifetime particle to traverse enough layers to be reconstructible. The disappearing track search improved sensitivity from a partial Run-2 to full Run-2 search without a major change in analysis strategy, which shows up as a vertical drop in Figure~\ref{fig:axe_plot}. Relative to searches for compressed sleptons, the search for disappearing-track higgsinos both has to (but also can) operate at a lower $A\times\varepsilon$, due to the larger higgsinos cross-section.

\subsubsection{Sleptons via Photon Production} Taken together, the rareness of sleptons (especially \slepton[R]) and the challenging signatures for $\Delta m_{\slepton} \lesssim m_{W}$ is a potent combination, and the LHC does not yet have sensitivity beyond LEP's reach for $20 \lesssim \Delta m_{\smuon[R]} \lesssim 80$~GeV, $10\lesssim\Delta m_{\selectron[R]}\lesssim~70$~GeV, and $2\lesssim\Delta m_{\stau[R]}\lesssim~60$~GeV~\cite{ATLAS:2025evx, ATLAS:2024fub}. Including LEP, there are even more striking gaps: there is no exclusion at all for \smuon[R] with $m_{\smuon[R]} > 40$~GeV in models with $0.1\lesssim\Delta m_{\smuon[R]} \lesssim 1.5$~GeV. 

However, the recent advances both in slepton sensitivity in the compressed corridor, as well as for the higgsino case discussed in Section~\ref{sec:higgsinos}, are clear indications that the full potential of the LHC is not yet exhausted, never mind the increased dataset expected from the HL-LHC. First, given that the superpartners of the neutrinos, $\tilde{\nu}$, are expected to have masses roughly in line with the $\slepton$~\cite{primer}, searches can target $\tilde{\nu}\slepton$ production as well as $\slepton\slepton$ production in a slightly less-simplified model scheme. Second, creative suggestions to improve sensitivity to $\slepton\slepton$ production at the LHC propose to take advantage of the fact that since \slepton[R] and \slepton[L] have the same electric charge, their production cross-sections via photons are the same.

At a hadron collider, photon fusion can uniquely occur in exclusive or ultra-peripheral collisions (UPC) in which the interacting hadrons -- either protons or heavy ions -- remain intact. As explored in Ref.~\cite{Beresford:2018pbt}, the ATLAS Forward Proton and the CMS–TOTEM Precision Proton Spectrometer detectors, which came online respectively in 2017 and 2016, allow measurement of the full missing momentum (\me) for proton-proton UPC by reconstructing the momentum of the outgoing protons. Tagging UPC events and reconstructing \me\ is a promising avenue for more deeply exploring compressed slepton production. The power of photon fusion production was highlighted by the observation of $\gamma\gamma\rightarrow WW$ production in proton-proton collisions, where the UPC collision is identified by reconstructing a primary vertex with no associated tracks except the $W$ decay products~\cite{ATLAS:2020iwi}.

Moreover, UPC $\gamma\gamma$ production can be enhanced for light particles in heavy ion collisions, due to the large electric charge of the ions, as noted in Section~\ref{sec:higgsinos}. Recent advancements in low $p_T$ lepton triggers using the ATLAS Transition Radiation Tracker FastOR Trigger suggest that lepton triggers with thresholds down to 500~MeV~\cite{trtFastOR} are possible in heavy-ion UPC events, which has the potential to dramatically increase the trigger efficiency, and therefore analysis sensitivity, for light, compressed sleptons\footnote{Because of a steeply falling interaction energy distribution in heavy-ion UPC events, production of heavy particles is heavily suppressed beyond around 200 GeV at the LHC for lead nuclei. Datasets with lighter nuclei have a raised energy cutoff and may provide unique opportunities.}. Between exclusive $\gamma\gamma$ production in proton collisions and UPC $\gamma\gamma$ production in heavy-ion collisions with low $p_T$ triggers, photon production of sleptons is an exciting future direction for searches at the LHC and HL-LHC.

\begin{marginnote}[]
\entry{Exclusive or UPC $\gamma\gamma$ production}{In exclusive or ultra-peripheral photon-photon production, the interacting protons or heavy-ions remain intact. The full momentum of the scattered protons can be measured with forward detectors, allowing reconstruction of \me. Additionally, exclusive or UPC $\gamma\gamma$ production involves no hadronic underlying event, producing a much more experimentally quiet collision than usually observed at a hadron collider.}
\end{marginnote}

\subsection{$R$-Parity Violation} 
\label{sec:rpv}

An additional $\mathds{Z}_2$ symmetry is commonly added to SUSY models such that $R$-parity $P_R$ is conserved, where 

\begin{equation}
P_R\equiv(-1)^{3(B-L)+2s}
\end{equation}

\noindent is $+1$ for SM particles and $-1$ for SUSY partners. $B$ and $L$ represent a particle's baryon and lepton number, respectively, and $s$ denotes its spin. The imposed $R$-parity conservation (RPC), where the total $R$-parity ($\prod_i P_{R,i}$) of the initial state is equal to that of the final state, is often justified by the \emph{accidental} $B$ and $L$ symmetries of the SM. RPC would imply that every vertex has an even number of sparticles. This then further implies that RPC models predict:

\begin{marginnote}[]
\entry{Accidental Symmetry}{An effective symmetry whose violation is not forbidden but would have to occur through non-renormalizable operators}
\end{marginnote}

\begin{itemize}
    \item sparticles must be produced in pairs at particle colliders, since the SM initial state necessarily has total $P_R=+1$,
    \item any sparticle decay must include a sparticle,
    \item and that the LSP is stable.
\end{itemize}

The assumption of RPC is often motivated by the SM's vertex-level conservation of $B$ and $L$, protects the proton from decaying, and provides a viable dark matter WIMP if the LSP is electrically and color neutral, such as for the lightest neutralino $\chi_1^0$.

RPC is not a particularly natural constraint on supersymmetric models; when building the most general MSSM Lagrangian from all renormalizable terms, there are tree-level violations of $R$-parity. RPC is frequently imposed as an ad-hoc additional symmetry in order to forbid these terms; couplings are set to zero by the model builder. In this way, RPC SUSY is not ``minimal''. A minimally-constructed supersymmetric extension would allow these terms.\footnote{However, the assumption of RPC need not be fully ad-hoc, as the symmetry may itself be accidental from an underlying true symmetry, as motivated in some compactified dimension string theories. The stringy mechanisms for creating an accidental $R$-symmetry may even be closely tied with the creation of an accidental Peccei-Quinn symmetry resolving the strong CP problem, discussed later in this review.~\cite{Baer:2025srs}} While RPC has many convenient features, it's reasonable to investigate its initial motivation. The $B$ and $L$ symmetries of the SM that often justify RPC are crucially not fundamental symmetries of the SM, and are instead accidental. We remain grateful that these accidental symmetries exist to protect the proton from decay, but proton decay alone does not necessitate strict conservation of $B$, $L$, and/or $P_R$; approximate conservation of $B$ or $L$ alone, or even only in light flavors, can suffice\footnote{In fact, in the general MSSM, dimension five operators appear that lead to rapid proton decay but don't violate RPC. These terms are also forbidden; even if we manually impose $B$ and $L$ conservation, these terms need to be separately suppressed to protect the proton from decay~\cite{Chen:2012tia}.}. If non-zero, allowed RPV couplings and combinations thereof are highly constrained by low energy limits on proton decay and neutron-antineutron ($n-\bar{n}$) oscillation, particularly for couplings to light flavor~\cite{Dreiner:1997uz}.

The structure of the general RPV MSSM Lagrangian is generated by its superpotential, which contains the following terms involving the supermultiplets containing both the sparticle and SM particle states:

\begin{equation}
W_{\rm RPV}=\mu_i L_i H_u + \frac{1}{2}\lambda_{ijk} L_i L_j\bar e_k +\lambda'_{ijk}L_iQ_j\bar d_k +\frac{1}{2}\lambda{''}_{ijk}\bar u_i\bar d_j\bar d_k,
\end{equation}
where $\mu_i, \lambda_{ijk}, \lambda'_{ijk}, \lambda''_{ijk}$ are the coefficients for the RPV interactions. The indices $i$, $j$, and $k$ denote the flavor generation of the superfield such that coupling coefficients can have flavor structure. Each of the trilinear couplings $\lambda_{ijk}$, $\lambda'_{ijk}$, and $\lambda''_{ijk}$ are a rank-3 tensor in flavor space, where $\lambda_{ijk}$ and $\lambda''_{ijk}$ are antisymmetric. On account of this anti-symmetry in the flavor structure, operators with repeated fields are zero, \emph{i.e.} $\lambda_{iik}=\lambda''_{ijj}=0$.

$L_i$ denotes the $i$-th generation superfield containing left-handed leptons and their superpartners; $Q_j$ is similar for the left handed quarks. $H_u$ represents the up-type Higgs doublet superfield. The lowercase symbols $e,u$, and $d$ denote left conjugate superfields containing the charged leptons, up-type quarks, and down-type quarks, respectively. For example, the ``UDD'' term, which is proportional to $\lambda''$, leads to vertices that can couple a squark and two quarks, and $\lambda''_{112}$ in particular could couple an up squark to a down and strange SM quark\footnote{Alternatively, for vertices involving the $\lambda''_{112}$ coupling, the squark line could come from either the first generation or second generation down-type superfields.}.

The RPV terms proportional to $\mu$, $\lambda$, and $\lambda'$ give rise to interactions that violate $L$ by one unit, and similarly the $\lambda''$ gives interactions that violate $B$ by one.
Compared to the RPC MSSM, the RPV MSSM superpotential contains 48 additional complex parameters, creating a rich phenomenology and introducing new theoretical and experimental challenges. For a recent thorough review of RPV collider phenomenology, please see Ref.~\cite{Dreiner:2023bvs}.

The magnitude of potential RPV couplings has a significant effect on collider phenomenology. If RPV couplings are comparable to the SUSY gauge couplings, the sparticles can avoid the traditional SUSY cascade topology, with heavy sparticles instead decaying directly to SM final states. If large enough, RPV couplings can cause single-sparticle production to occur at large rates. However, large RPV couplings face significant low- and high-energy experimental constraints~\cite{Barbier:2004ez}. Therefore, many arguments may suggest that RPV couplings should be small, even if non-zero. If such couplings contribute significantly less than the gauge couplings, SUSY cascade decays to the LSP are likely to still dominate, where the LSP then decays to an all-SM final state via an RPV coupling. In such models, the traditional neutralino WIMP DM paradigm fails as the \N{1} is no longer stable\footnote{In such scenarios, the DM content might still live elsewhere in the sparticle spectrum with contributions from long-lived gravitinos, discussed in Sect.~\ref{sec:gravitinos} or axions and axinos, discussed in Sect.~\ref{sec:axion}, among others.}. Moreover, the traditional \met-based SUSY search program becomes less relevant.

\subsubsection{RPV decays with leptons} The $L$-violating terms can result in decays involving neutrinos. While not optimized for them, \met-based RPC searches can retain some sensitivity to such models. While much of the RPV search program focuses on the trilinear terms proportional to $\lambda_{ijk}$, $\lambda'_{ijk}$, and  $\lambda''_{ijk}$, the bilinear term proportional to $\mu_i$ can give rise to signatures not often seen elsewhere in SUSY searches. Several searches have included this coupling~\cite{ATLAS-CONF-2015-018, ATLAS:2023lfr}. Additionally, ATLAS has performed searches for tri-lepton resonances in final states without \met\ specifically targeting this scenario~\cite{ATLAS:2020uer}. These searches target chargino production with bilinear RPV decays to a lepton and Z boson. While the resulting tri-lepton resonance signature is striking, this coupling can give rise to resonant lepton + vector boson signatures generally.

\begin{marginnote}[]
\entry{Lepton + $V$ Resonance}{RPV Electroweakino decays can resonantly produce a lepton and vector boson, which can lead to a resonant tri-lepton system.}
\end{marginnote}

\subsubsection{Hadronic RPV decays} In contrast, signatures without \met\ can pose additional experimental challenges. The baryonic RPV term proportional to $\lambda''$, in particular, can give final states with large jet multiplicities with no \met. As a hadron collider, the LHC produces copious hadronic activity that could easily hide such supersymmetric signatures under \textbf{large backgrounds}. 

\begin{marginnote}[]
\entry{Multi-jet Signals}{Sparticle decays to fully hadronic final states with no \met. While there may be internal resonant structures, combinatorial and SM multi-jet backgrounds make searches challenging.}
\end{marginnote}

Beyond a large QCD multi-jet background, the multi-jet signatures that naturally arise with nonzero $\lambda''$ are experimentally complicated by the signal topology. For light jet signatures with no experimental flavor handles, there arises a \textbf{signal combinatorial problem}. If pair-produced sparticles decay to many (4-10) indistinguishable hadronic jets, using multi-object invariant mass scales becomes highly sensitive to the correct groupings of jets. As the jet multiplicity and complexity of the decay chain increases, this combinatorial topology becomes difficult to correctly resolve, which reduces the rejection power against SM background. In such SUSY models, the internal resonance masses are yet unknown, such that the rich kinematic information that is in principle available is difficult to extract. In the case that rich kinematic information is buried in a high dimensional feature space, modern machine learning methods should be a sharp tool, as studied in Refs.~\cite{Badea:2023jdb} and \cite{Badea:2022dzb}. The LHC experiments have attempted to solve this problem in a number of ways~\cite{ATLAS:2024kqk,ATLAS:2017jnp,CMS:2024ldy,CMS:2022usq} and some have decided to integrate over this problem entirely~\cite{ATLAS:2015xmt,ATLAS:2024kqk}, leading to moderate sensitivity, but are ultimately limited by the combination of ambiguous signal combinatorics and a large SM background.

\subsubsection{Long-lived RPV signatures} The collider phenomenology varies greatly across the RPV coupling space with the potential for hierarchy introduced in the flavor structure. Ensuring comprehensive search coverage requires sensitivity to be evaluated across a broad range of signatures. For a simplified model of a single nonzero RPV coupling, very small couplings can lead to LSP decays far outside of the detectors, resulting in sensitivity from the canonical \met-based SUSY program. As RPV couplings increase, the lifetime of the LSP decreases, leading to displaced decays where searches for LLPs will maximize sensitivity. This wide array of signatures, comprising \met-based, LLP, no-\met, and single production signatures, is further complicated by the transition regions between them and more complex (\emph{i.e.} less contrived) flavor structures with an ensemble of active RPV couplings.

Very small RPV couplings can be theoretically motivated. In dynamical RPV~\cite{Csaki:2013jza}, the violation is generated dynamically in a hidden sector with some messenger scale $M$ and this symmetry breaking is closely tied with the SUSY breaking mechanism. In this model, the RPV interactions are highly suppressed as $\nicefrac{1}{M^2}$, justifying these dimensionless couplings $\ll\mathcal{O}(1)$. As mentioned above and discussed further in Section~\ref{sec:minisplit}, sensitivity for very-small RPV couplings is dominated by searches for LLPs, as long as the LSP is either charged or decays within the detector. RPV SUSY can uniquely give long-lived metastable LSPs and can accommodate nearly any sparticle as the LSP.

If long-lived higgsinos are produced with displaced decays via the UDD coupling, we again have a multi-jet signature, but one that is now strikingly different from any SM background. Such a scenario can give reconstructed displaced vertices from the charged particle tracks from these jets. This process is rare because of the low higgsino production cross-section, but is also challenging because of the special reconstruction. However, once this effort is in place, the signature is so striking that cross-section limits from such analyses are unmatched at the LHC as discussed for Figure~\ref{fig:axe_plot}. At low higgsino masses, however, the trigger systems only see a low energy hadronic final state, exactly the signature LHC triggers are designed to reject; low mass all-hadronic searches suffer from trigger limitations, displaced or not.

\begin{marginnote}[]
\entry{Displaced Vertex + Jets}{
Reconstructed charged particle tracks are fit to a common vertex measurably displaced from the primary collision vertex, particularly in the transverse direction.}
\end{marginnote}

\subsubsection{Comprehensive RPV Coverage} 
\begin{figure}
\includegraphics[width=6.0in]{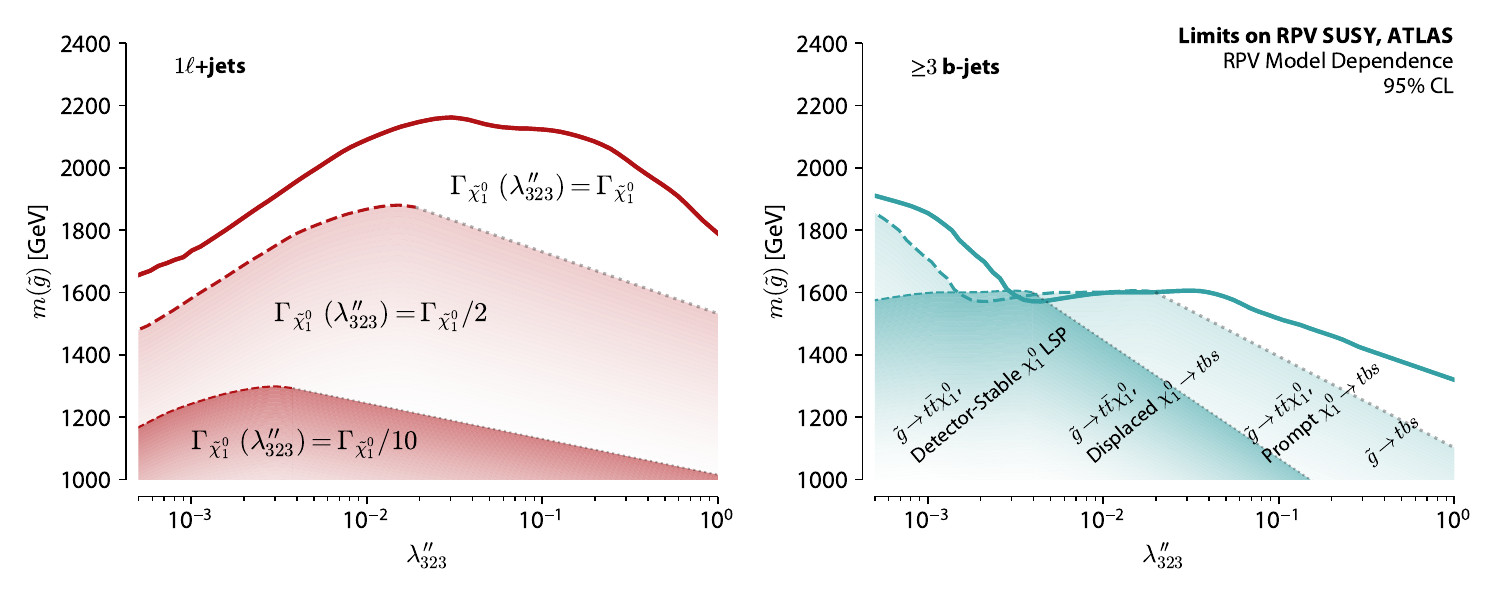}
\caption{Solid lines show the ATLAS constraints on the $\tilde{g}$ mass in RPV models with only a nonzero $\lambda''_{323}$ as a function of the coupling strength~\cite{ATLAS-CONF-2018-003}. With varied coupling, the sensitivities of the $1\ell$+jets channel (red, left) and $\ge3$ b-jets channel (blue, right) vary significantly. Relaxing the assumption that a singular RPV coupling is nonzero forces the partial width of the $\tilde{\chi}_1^0$ due to $\lambda''_{323}$ to be a fraction of the total width. The dashed curves show the reduction in sensitivity if $\lambda''_{323}$ is responsible for only half or 10\% of the total width. The gray dotted lines represent an unevaluated interpolation between the end of the dashed curves and the large coupling limit deduced from branching ratio scaling.
}
\label{rpv}
\end{figure}

In Ref.~\cite{ATLAS-CONF-2018-003}, ATLAS performed a reinterpretation of many analyses, making efficient use of containerization and automation in analysis software. The collaboration set limits as a function of a single RPV coupling strength, focusing on several $\lambda''$ couplings and assuming several separate sparticle spectra. This exercise performed by ATLAS unveiled several gaps in coverage. In Figure~\ref{rpv}, exclusion contours from ATLAS show the 95$\%$ CL lower limit on $m_{\tilde{g}}$ as a function of the $\lambda''_{323}$, which is assumed to be the only nonzero RPV coupling. This assumption is of course overly strict; any reasonable model of RPV would likely have some flavor structure in the coupling space~\cite{Csaki:2011ge,Csaki:2013jza}\footnote{In Run-1, ATLAS performed systematic scans over the flavor structure of the three trilinear RPV terms~\cite{ATLAS:2015xmt,ATLAS-CONF-2015-018}, but without accounting for changing diagram topology at particularly small or large coupling values.}. 
We test the robustness of these exclusions by assuming that LSP decays via $\lambda''_{323}$ contribute only a portion of the total width. As the fraction of the width given by this coupling decreases, the limits recede quickly, with some mass limits reduced to below $1$~TeV.

This study demonstrates that, as always, reliance on overly simplified models can lead to a misleading picture of coverage. In models with a very \textbf{large parameter space}, ensuring robust coverage requires dedicated, organized investigations in order to cover significant regions of parameter space and to discover gaps in experimental coverage. This program is challenging, not because any particular signature is intrinsically more challenging than the rest of the LHC search program, but because of the \textbf{rapid signature variation} over a very \textbf{large parameter space}.

The exercise of measuring the complex interplay between multiple signatures for varied RPV coupling strength is a major step away from traditional simplified SUSY models that became the norm during Run-1 of the LHC to a more model-driven interplay, one level zoomed out from focusing on a single diagram. If a traditional simplified model evaluates a single diagram across a sparticle mass space, the addition of varied coupling strengths amounts to a \emph{de-}simplification of the model. Examining parameter spaces of even further \emph{de-simplified models}, for example by exploring the flavor structure of the theory, can easily lead to the identification of further low-mass gaps in sensitivity and should be prioritized by the community.

\section{EXTENDED-MSSM SIGNATURES}

Extensions of the MSSM designed to address other issues in the SM or new problems introduced in the MSSM will lead to additional particle or coupling content, likely adding complexity to the search space. Expanding the dimensionality of the model parameter space makes effective exploration challenging, as discussed above.

Many of the examples below do not necessarily constitute rare processes, but easily result in challenging signatures, as many were designed or popularized in an era of existing LHC constraints. The discussion below is far from a complete overview of extended MSSM signatures, but attempts to highlight examples with particular challenges and/or dedicated LHC search programs.

\subsection{Gravitinos}
\label{sec:gravitinos}

The gravitino, $\tilde{G}$, is the superpartner to the not-particularly-SM graviton. The graviton, as a quantum excitation of the gravitational field for which we have only recently measured classical gravitational waves, is expected to be extremely difficult to detect. It is expected to be massless to ensure long-range gravity, and its couplings with SM particles are expected to go like $1/M_P$, \textit{i.e.}, very weak. Conversely, the gravitino may be easier to discover than the graviton, and we may see it (or direct evidence of it) first. The $\tilde{G}$ differs in several important ways from its partner. First, the $\tilde{G}$ can acquire a mass from SUSY-breaking effects. It may be light ($\lesssim O(m_W)$ in supergravity models~\cite{ELLIS198499} and extremely light (possibly $O(1)$~eV, depending on the SUSY-breaking scale) in Gauge-Mediated SUSY-Breaking (GMSB) models~\cite{Giudice:1998bp}, but not massless. Second, its couplings depend on the SUSY-breaking scale, and may be significantly larger than those of the graviton (although still weak)~\cite{Luty:1998np}. Finally, its interactions with the other MSSM particles may lead to very different phenomenological effects than the graviton. Many GMSB models predict that the $\tilde{G}$ is the LSP~\cite{Giudice:1998bp}; in this case, if other sparticles can be kinematically produced at colliders, their decay chain will usually end with a $\tilde{G}$. 

While a minimal GMSB model makes fairly strict predictions about the relationship between sparticle masses, it is hard to reconcile its predictions with a 125~GeV Higgs boson and with the existing squark and gluino limits from the LHC. In response, a more flexible General Gauge Mediation (GGM) class of models is still consistent with experimental results~\cite{Meade:2008wd}. Many gauge mediation models share common predictions, including flavor universality among the sfermion masses and a $\tilde{G}$ LSP\cite{Meade:2008wd}. As the $\tilde{G}$ couples very weakly to other sparticles, it is also a common feature of GGM models that the NSLP acquires a macroscopic lifetime.

\begin{marginnote}[]
\entry{LLP Direct Detection}{Searches which target the direct interaction of a charged LLP with the detector, usually with techniques that are sensitive to the particle's velocity to distinguish a heavy LLP from lighter SM LLPs. To first-order, direct detection searches are insensitive to the LLP decay mode, although second-order effects like isolation or \met\ can change with the decay mode.}
\end{marginnote}

\begin{marginnote}[]
\entry{LLP Indirect Detection}{Searches which target the SM decay products of an BSM LLP. The decay products are tagged as displaced in time or in space and usually target a specific decay mode.}
\end{marginnote}

If the NLSP is a slepton, which is a common although not universal feature of GGM models~\cite{Meade:2008wd}, slepton pair-production would produce distinctive experimental signatures. The optimal LLP search techniques depends on the $\slepton$ lifetime, $\tau_{\slepton}$, changing from \textit{direct detection} for $\tau_{\slepton} \gtrsim 1 $~ns to \textit{indirect detection} at shorter $\tau_{\slepton}$. The displaced lepton searches dominate at short $\tau_{\slepton}$ and the disappearing track signature reaches peak sensitivity at $\tau_{\slepton} \approx 1$~ns. For any longer lifetimes, searches for tracks with anomalous ionization and/or anomalous time-of-flight (ToF) measurements are most sensitive. See Figure~\ref{fig:staus} for an overview of the sensitivity to $\stau\rightarrow \tau \tilde{G}$, as a function of $m_{\stau}$ and $\tau_{\slepton}$.

The ATLAS and CMS experiments have targeted the anomalous ionization and time-of-flight signatures since early Run 1~\cite{ATLAS:2012hja, ATLAS:2011ghv, CMS:2012wcg}, although interpretations for sleptons had to wait until enough luminosity had been collected. To measure ionization, a detector must record the charge deposited in an individual layer. At the LHC, the inner detector subsystems and the muon systems can do so. To usefully measure ToF for SUSY signatures, the subsystem must have a timing resolution of less than $1$~ns, which is achievable with the calorimeters (although this is energy dependent) and muon systems. Combining multiple discriminants from different subdetectors improves background rejection, though at the price of signal efficiency, and, if the LLP must travel to the calorimeters and/or muon system, a loss of signal acceptance for shorter lifetimes. To reconstruct and calibrate the ionization and timing measurements requires a custom program which faces noticeable computing constraints in accessing the low-level detector information needed for precise measurements. Additionally, the measurement of ionization in the innermost detector layers, \textit{i.e.} the Pixel detector, faces extreme effects from radiation damage and run-by-run corrections are required~\cite{ATLAS:2022pib}. The current sensitivity to long-lived sleptons from these searches extends to nearly 600~GeV at $\tau_{\slepton} \approx 10$~ns ~\cite{ATLAS:2025fdm} and up to 700~GeV for $\tau_{\slepton} \gtrsim 100$~ns~\cite{CMS:2024nhn}. These results assume $m_{\slepton[R]} = m_{\slepton[L]}$ but are largely independent of the slepton flavor.

\begin{marginnote}[]
\entry{Anomalous ionization}{Searches which target charged LLPs by reconstructing a track with ionization energy loss, $dE/dx$, which is larger than that expected for a minimum ionizing particle with the same momentum. As $dE/dx$ depends on $\beta\gamma$, the LLP mass can be reconstructed with $\beta\gamma = \frac{p}{m}$.}
\end{marginnote}

\begin{marginnote}[]
\entry{Time-of-Flight}{Searches which target charged LLPs by reconstructing a track which arrives at a specific detector subsystem later than expected for a particle with $\beta = c$.}
\end{marginnote}

For 10~ps~$\lesssim \tau_{\slepton} \lesssim 1$~ns, the indirect displaced lepton signature dominates the sensitivity. Note an important phenomenological difference between indirect and direct detection searches: while direct detection searches are, to first-order, insensitive to $\Delta m_{\slepton}$, indirect searches require a minimum $\Delta m_{\slepton}$. Therefore, while the disappearing track, anomalous ionization, and ToF searches are useful for the co-annihilation slepton models discussed in Section~\ref{sec:sleptons}, displaced lepton searches are not particularly relevant to RPC MSSM scenarios without GGM, as there is no mechanism in these models to give sleptons a long lifetime in scenarios with large $\Delta m_{\slepton}$. Displaced leptons searches are useful in models beyond GGM, however; as discussed in Ref.~\cite{Evans:2016zau}, displaced leptons also show up in RPV models with LLE couplings, as well as in freeze-in dark matter models~\cite{Das:2025oww}. Unlike GGM models, both RPV and dark matter models can predict signatures of displaced leptons with differing flavors. 

In 2016, it was pointed out that the LHC had a notable gap in slepton coverage, as the only displaced lepton searches at that time targeted different flavor leptons~\cite{Evans:2016zau}. The ATLAS experiment soon followed with a search targeting same-flavor displaced lepton signatures~\cite{ATLAS:2020wjh}, bringing peak sensitivity to sleptons with $\tau_{\slepton} \approx 100$~ps, of up to 700~GeV for $m_{\smuon[R]} = m_{\smuon[L]}$ and  $m_{\selectron[R]} = m_{\selectron[L]}$, and up to 300~GeV for $m_{\stau[R]} = m_{\stau[L]}$. The search relies on custom tracking for large-radius tracks resulting from displaced decays with large impact parameters~\cite{ATLAS:2017zsd}. In addition to the \textbf{unusual reconstruction}, the search had to re-optimize the lepton identification criteria for displaced leptons. The custom tracking increases the efficiency for reconstructing leptons with impact parameters greater than 20~mm by more than an order of magnitude, and the custom identification algorithms improve the efficiency again by a factor of two or more~\cite{ATLAS:2020wjh}.

\begin{marginnote}[]
\entry{Displaced Leptons}{An event with one or more leptons which are displaced in time or space from the collision point. For slepton production in GGM models, the two displaced leptons in an event do not share a common vertex.}
\end{marginnote}

Searches for long-lived sleptons in the plane of Figure~\ref{fig:axe_plot} highlight how different LLP searches can be from prompt searches, and how different they can be from each other, depending on the details of the lifetime, signature, and model parameters. While searching for tracks with anomalous ionization or ToF requires \textbf{unusual reconstruction} and custom calibrations that are inherently challenging, the reward is a signature with little background, which can operate at a relatively high $\axe \approx 60\%$ for sleptons with detector-stable lifetimes ($\tau_{\slepton} \gtrsim 100$~ns), which is not possible for the prompt searches. While detector stable track-based searches can use the muon system to trigger, this is inefficient for lifetimes of $\tau_{\slepton} \lesssim 10$~ns, which rely on a \met\ trigger, and therefore operate at lower values of $\axe \approx 5\%$. Moreover, the ATLAS and CMS anomalous ionization searches operate at different $\axe$ working points not just due to different triggers and analysis choices, but also because of differences in the number of layers that can be used to measure ionization in the different detectors and the corresponding background-rejection power of the $dE/dx$ measurement~\cite{CMS:2024nhn, ATLAS:2025fdm}. Nonetheless, they reach similar cross-section sensitivities. 

The displaced slepton results tell a similar story. CMS tends to work at slightly higher \axe\ values than ATLAS where the backgrounds are larger, but overall, the two experiments achieve similar cross-section sensitivities for the same lifetime while operating with different optimizations. Note that the first ATLAS displaced lepton analysis optimized for $\tau_{\slepton} = 0.1$~ns and lost sensitivity quickly for $\tau_{\slepton} = 0.01$~ns due to tight impact parameter requirements. An additional search optimized for shorter lifetimes, with an order of magnitude improvement in sensitivity~\cite{ATLAS:2023ios}, highlighting both the challenges and opportunities in effectively scanning a \textbf{large parameter space} with \textbf{rapidly changing signatures}. 

In GMSB and GGM models, $\slepton$ are not the only viable NSLP; $\tilde{B}$ are also a common possibility~\cite{Meade:2008wd}. In this case, pair-produced electroweakinos can each decay to a long-lived $\N{1}$, which in turn decays to a SM boson and $\tilde{G}$. This process can produce a distinctive non-prompt photon signature. The boson in the $\N{1}$ decay can be either a photon directly, or a $H$ or $Z$ which in turn decays to a pair of photons or electrons (which, if displaced, can be identified as photons). ATLAS and CMS have used the precise pointing and timing capabilities of the electromagnetic calorimeters to target GGM motivated-models in the case of pairs of individual non-prompt photon production~\cite{ATLAS:2022vhr, CMS:2019zxa}, as well as in the case that the two reconstructed displaced photons share a common vertex from a $H$ or $Z$ decay~\cite{ATLAS:2023meo}.

\begin{marginnote}[]
\entry{Non-prompt photon}{A photon which originates from a displaced decay of a BSM LLP. The photon is tagged as non-prompt from its delayed time of arrival in the calorimeters or non-pointing due to its reconstructed orientation}
\end{marginnote}

Looking forward, the collaborations have identified opportunities to improve searches for unconventional signatures by designing custom triggers. Numerous new triggers targeting long-lived signatures were implemented for Run 3 by both ATLAS and CMS~\cite{ATLAS:2024xna, CMS-DP-2023-043}. A search for displaced leptons using a custom trigger, an improved version of the displaced tracking algorithm~\cite{ATLAS:2023nze}, and improved displaced electron identification was conducted with the first two years of Run 3 data by the ATLAS experiment~\cite{ATLAS:2024vnc}. Together, the improvements expanded the lifetime range of the search from $\tau_{\slepton} \approx 10$~ns to $\tau_{\slepton} \approx 40$~ns at $m_{\slepton} \approx 200$~GeV for $\smuon$ and $\selectron$ production, highlighting the potential impact of trigger and reconstruction improvements in Run 3, and beyond. The upgraded detector and trigger systems for the HL-LHC era will continue to provide room for creative improvements for unconventional and challenging signatures.

\subsection{Stealth SUSY}

Stealth SUSY~\cite{Fan:2011yu} proposes a mechanism for exclusively producing low-\met\ SUSY signatures while preserving RPC. Introducing only a small amount of SUSY-breaking results in small mass splittings between the fermion and boson partners. New particles are then added on top of the MSSM particle content: a new scalar singlet $X$ and its fermionic SUSY partner singlino $\tilde{X}$. The mass splitting between these states remains small and the $\tilde{X}$ can be the NLSP, decaying to $X$ and the LSP. Due to the small $X-\tilde{X}$ mass splitting, there is limited phase space for the LSP and low-\met\ signatures become natural. For small enough mass splittings and/or large enough SUSY-breaking scales, singlinos can become long-lived.

Experimental searches for low-MET searches for compressed spectra, as discussed in Sect.~\ref{sec:compressed}, generically have sensitivity to Stealth SUSY. However, compressed spectrum searches often target low intrinsic \met\ signatures with low momentum visible objects. As Stealth SUSY decay chains are marked by compression in the production of the LSP, high momentum particles are easily produced in the traditional sparticle cascade, making the signature more akin to the RPV signature space. Without explicit \met\ requirements, many RPV searches discussed in Sect.~\ref{sec:rpv} can have sensitivity. In addition to experimental reinterpretations~\cite{ATLAS:2020viz}, multiple dedicated searches for Stealth SUSY have been performed by the LHC collaborations~\cite{CMS:2023zuu,ATLAS:2018tup,CMS:2025bxo,CMS:2019uzi}.

\begin{marginnote}[]
\entry{Low \met, high $p_T$ objects}{Small mass splittings at the end of a Stealth SUSY decay chain give low intrinsic (but promotable) \met\ produced in association with a cascade of high momentum objects.}
\end{marginnote}

\subsection{Supersymmetric Axion Models}
\label{sec:axion}

The apparent lack of CP violation in the strong force, known as the Strong CP Problem, can be resolved by an underlying Peccei-Quinn (PQ) symmetry and a new pseudoscalar particle, the \emph{axion}. If it exists, the axion may contribute to the observed Dark Matter density.
In BSM extensions that include supersymmetry and a PQ symmetry, the axion is included in a new supermultiplet containing a scalar \emph{saxion} particle $s$ and a fermionic \emph{axino} $\tilde{a}$, which may also contribute to the DM.~\cite{theoristSUSY_review}

While the LHC is not well suited to search for the QCD axion, the supersymmetric addition of these new states makes for a complex set of collider signatures, depending on the masses of all three new particles. The standard SUSY decay chains can be longer with some models motivating a low-mass axino LSP dark matter candidate. For example, with an axino LSP, cascade decays may be extended by the NSLP decay to an axino and a photon~\cite{Choi:2008qh}. Since the new multiplet's couplings are suppressed by a large PQ scale, long-lived NLSPs naturally arise. Collider searches for long-lived particles targeting proper lifetimes of 10~ps to 100~ns can have sensitivity to such axion-sector-induced lifetimes consistent with Big Bang Nucleosynthesis constraints. 

Two major QCD axion solutions exist: the KSVZ model where the axion primarily interacts with gluons and gains loop-level photon couplings, and DFSZ where the axion couples to SM fermions at tree level and an additional Higgs doublet is added~\cite{Kim:2008hd}. Both models can accommodate SUSY extensions, but SUSY DFSZ has several notable features. As both the minimal SUSY extensions and DFSZ models require two Higgs doublets, SUSY DFSZ is a natural fit. The new particle content of SUSY DFSZ creates an anomaly cancellation that suppresses the coupling to photons, rendering cavity photon-based experiments less sensitive by an order of magnitude~\cite{Bae:2017hlp}. On the other hand, introducing a DFSZ-style PQ symmetry helps resolve the SUSY $\mu$ problem, where the MSSM superpotential contains a term $\mu H_{u} H_{d}$ pushing the Higgsino mass parameter $\mu$  to the Planck scale by coupling it to the two Higgs doublet superfields~\cite{Kim:1983dt}. This term is forbidden in DFSZ extensions and is replaced with a term suppressed by the Planck scale to be near the weak scale~\cite{theoristSUSY_review}. The \textit{effective} $\mu$ is set by the PQ scale. This symbiotic relationship allows DFSZ axions to further evade cavity experiments \textit{and} frees the SUSY model from the $\mu$ problem.\footnote{In the DFSZ framework, the Higgs doublets carry PQ charge. As such, in SUSY DFSZ, the Higgsinos are also carrying this charge which in principle allows the axino to mix (at very low rates) and contribute to a total of five neutralino eigenstates instead of the traditional four in the MSSM.}

A SUSY DFSZ model can simultaneously evade cavity-based direct detection, populate the DM relic density with a rich family of stable dark states, and provide challenging collider signatures. The RPC SUSY program at the LHC can retain sensitivity to SUSY cascades that end in a stable axino giving \met, but the coupling of the axion to the SM particles determines both the axion mass and the NLSP decay width to the axino, such that the lifetime of the NLSP scales as $1/m_{a}^2$, reaching the mm scale and beyond for a wide range of models. Much like the RPV case described above, the SUSY DFSZ phenomenology can range from reduced-\met\ prompt signatures for large axino couplings to LLP signatures to RPC-like MET-based signatures. Additional signatures can be explored in models where the gravitino is the LSP and the axino is the NLSP.

While numerous searches for axions (and axion-like particles, or ALPs) outside of the context of SUSY have occurred at the LHC~\cite{Hook:2019qoh}, few dedicated searches exist assuming a SUSY context, despite the rich phenomenology. (One such search is presented in Ref~\cite{ATLAS:2021yqv}.) This is likely due to much of the signature space mapping onto existing searches for SUSY and more generic signatures. Prospects for future sensitivity at the HL-LHC and details of the relevant phenomenology are described in Ref.~\cite{Curtin:2018mvb}.

In the program of modern collider LLP searches, particularly from the CMS and ATLAS Collaborations, many analyses focus on generic displaced signatures that are striking enough to allow for model-inclusive signal regions that have small backgrounds. Displaced \mbox{(di-)photon}~\cite{CMS:2025urb,CMS:2024vjn}, \mbox{(di-)lepton}~\cite{CMS:2022qej,CMS:2021sch}, and jet signatures~\cite{CMS:2024bvl,CMS:2022wjc,CMS:2021yhb} in principle have sensitivity to long-lived NLSP cascades to axino LSPs via displaced SM H/W/Z/$\gamma$ boson emission. These searches are ripe for reinterpretation, partially due to a Run-2 push for multi-dimensional efficiency maps, particularly from the ATLAS SUSY program, and have been reinterpreted in a SUSY Axion context in Ref.~\cite{Hoshino:2025lqz}.

\begin{marginnote}[]
\entry{Displaced boson decays}{The displaced decay of energetic SM bosons can give rise to displaced di-lepton, di-jet, and di-photon signatures, or even more complex, distinctive signatures.}
\end{marginnote}

\subsection{Mini-Split SUSY}
\label{sec:minisplit}

Rather than focusing on addressing the Higgs naturalness problem, an alternate approach to SUSY model-building focuses on the benefits of gauge unification, a WIMP dark matter candidate, and additional avenues for CP violation. In traditional SUSY models, the scalars and gauginos gain tree-level masses from the SUSY breaking scale via soft mass terms, requiring relatively low SUSY breaking scales to ensure sparticle content near the weak scale. In \emph{Split SUSY}, the scalar masses are allowed to grow with a large SUSY breaking scale at tree level, while the gaugino masses are assumed to be protected by some symmetry near the electroweak scale~\cite{Arkani-Hamed:2004zhs, Giudice:2004tc}. In contrast, in \emph{mini-split SUSY}, the gauginos are prevented from having tree-level masses from the SUSY breaking scale but retain a loop-level coupling to this high scale.  Different SUSY breaking models predict differing numbers of loops and magnitudes, but each additional loop suppresses the mass by a factor $1/(4\pi)^2$. While the scalars are driven to the SUSY breaking scale, the gauginos are 2 or 4 orders of magnitude lighter for 1- or 2-loop mass suppression, respectively.~\cite{Arvanitaki:2012ps}

Under a mini-split paradigm, a 125~GeV Higgs implies a SUSY breaking scale between $10^1-10^4$~TeV, with high $\tan\beta$ values favoring the lower end of that range. One-loop (two-loop) suppression of gaugino masses implies gauginos at the $100$~TeV ($1$~TeV) scale for $\tan\beta\approx1$. The assumption that the gauginos are prevented from tree-level coupling to the SUSY breaking scale predicts a specific, phenomenologically-rich region of model space. This mechanism and its relationship to the electroweak scale naturally generate LLP signatures at colliders.

In mini-Split SUSY, where gaugino masses are given by a two-loop coupling to the SUSY breaking scale, gauginos at the TeV scale will have suppressed widths as their decays involve highly off-shell scalars. A gluino at this scale will gain a long lifetime before decaying as $\tilde{g}\rightarrow q\bar{q}\chi_1^0$, $\tilde{g}\rightarrow q\bar{q}'\chi_1^\pm$, or $\tilde{g}\rightarrow g\chi_1^0$ via a squark loop. Its lifetime is parametrized as: 

\begin{equation}
 \Gamma^{-1}\sim\frac{4}{N}\left(\frac{m_S}{10^3~{\rm TeV}}\right)^4\left(\frac{1~{\rm TeV}}{m_{\tilde g}}\right)^5\times10^{-3}~\rm{ns},
\end{equation}

\noindent where $N$ is a model-dependent $\mathcal{O}(1)$ parameter, $m_S$ is the scalar mass near the SUSY breaking scale, and $m_{\tilde{g}}$ is the gluino mass. This lifetime is highly sensitive to both mass scales and easily produces experimental detector signatures from prompt decays to detector-stable gluinos, with a large range of intermediate displaced vertex signatures possible~\cite{Lee:2018pag}.

If the gluino gains a lifetime longer than the SM QCD hadronization timescale, as a color-carrying particle, it will form a colorless bound state with SM quarks and gluons known as an $R$-hadron\footnote{Despite the name, the formation of an $R$-hadron does not require any conservation of $R$-parity. The name uses $R$ to simply denote ``supersymmetric''.}. The added parton content and QCD mass will create a spectrum of possible bound states above the bare gluino mass. Depending on the composition of this $R$-hadron, it can be electrically neutral or charged, with possible doubly charged states. The relative production rate of these $R$-hadrons will depend on the spectrum structure, with the phenomenology being largely defined by the properties of the lowest mass $R$-hadron state. Understanding this spectrum is an exercise in pure SM QCD as it involves only SM QCD charge. The assumed spectrum has historically relied on bag models and constituent quark mass models~\cite{ATL-PHYS-PUB-2019-019}, but more sophisticated methods may be of interest to the lattice QCD community and is yet to be studied. Searches employing direct detection techniques introduced in Section~\ref{sec:gravitinos} may be highly sensitive to the spectrum structure and order, as these searches will tend to require that the $R$-hadron is charged.~\cite{Lee:2018pag,Bierlich:2022pfr,ATL-PHYS-PUB-2019-019} 

The propagation of these $R$-hadrons through the detector is a complex \textbf{simulation challenge} that breaks the standard collider simulation workflow ordering. Particle-level generators such as Pythia can create $R$-hadrons, but they must be propagated through the detector material and magnetic fields using dedicated specialized interaction models that can be interfaced to GEANT4~\cite{Mackeprang:2006gx,Mackeprang:2010swu}. If the visible decay of the long-lived gluino is to be simulated, a complex handoff is required, where the detector simulation in GEANT4 must be interfaced to a generator capable of complex gluino decays and a parton shower such as Pythia. Such an interface is highly experiment-specific with an implementation from ATLAS~\cite{ATL-PHYS-PUB-2019-019} and one under development from CMS.

Searches for gluino $R$-hadrons motivated by mini-Split SUSY have been performed since the beginning of LHC data and often exceed standard gluino searches in mass reach due to their striking LLP signatures. Starting in Run-2, the prompt SUSY search program sets limits on small lifetimes that are not large enough to lead to a complete loss of reconstruction efficiency~\cite{ATLAS-CONF-2018-003}. These interpretations are crucial for ensuring no gap in coverage exists.

\begin{marginnote}[]
\entry{Displaced Vertices + \met}{
A high-mass displaced vertex is reconstructed associated with significant \met. The \met can be used to trigger efficiently.}
\end{marginnote}

Many searches are performed using reconstructed secondary vertices from the decay of the $R$-hadron using charged particle tracks~\cite{CMS:2024trg,CMS:2020iwv,ATLAS:2017tny,ATLAS:2020xyo,ATLAS:2023oti} where displaced hadronic jets leave a distinctive signature if the \textit{Displaced Vertex} (DV) is successfully reconstructed. Reconstruction losses at the displaced track and subsequent vertexing steps often require specialized algorithms~\cite{ATL-PHYS-PUB-2019-013}, but for displaced high energy jets, the charged particle multiplicity is large enough to make the overall reconstruction efficiency large for proper lifetimes below $\mathcal{O}(1)$~ns where the ATLAS and CMS silicon trackers have coverage. This signature allows such searches to work in a very low background regime; in Figure~\ref{fig:axe_plot}, the DV searches all exclude cross sections as low as possible for a given signal $A\times\epsilon$ and dataset luminosity. Since these searches simultaneously reach very large $A\times\varepsilon$, they are able to probe some of the smallest cross sections available to the LHC, exceeding the sensitivity of more conventional searches, and in fact the rest of the LHC analysis program. 
In the search performed in Ref.~\cite{ATLAS:2023oti}, even improving $A\times\varepsilon$ beyond its roughly 60\% to the ideal 100\% could only improve the cross section limit by a factor of 2, and only if a zero background region were still possible. If an LLP with a tracker-compatible lifetime decays to jets, this search is nearly the ideal analysis.

Calorimeter signatures can also be used to find late or non-projective jets which do not point back to the collision point. CMS performed a full Run-2 search using calorimeter timing to identify late jets with sensitivity to slow moving, late decaying $R$-hadrons~\cite{CMS:2019qjk}. 

Direct detection searches look for interactions of the (charged) $R$-hadron with the detector before its decay, usually by looking for anomalous ionization signatures along reconstructed tracks. Slow moving, heavy, charged particles can produce measurably more ionization energy loss than a minimum-ionizing particle (MIP), which can be used to distinguish them from highly boosted SM particles~\cite{CMS:2024nhn,ATLAS:2022pib, ATLAS:2025fdm}. A search from ATLAS~\cite{ATLAS:2022pib} notably observed a $3.6\sigma$ local excess with ionization consistent with a roughly $2$~TeV gluino signature. However, time of flight measurements consistent with speeds $\beta=1$ were not consistent with this heavy particle interpretation. Ref.~\cite{Giudice:2022bpq} proposed that the combination of high ionization and high speed could instead be explained with highly boosted doubly-charged LLPs. The CMS Collaboration responded with a search that excluded the observed ATLAS excess cross section at 95\% CL even under the most favorable assumptions on the ATLAS selection $A\times\varepsilon$ for such a model\footnote{This exclusion holds as long as the boosted doubly charged LLP would fire the CMS muon trigger, reconstruction, and loose identification.}.

\begin{marginnote}[]
\entry{Stopped Particles}{Large energy depositions are observed highly out-of-time with the beam, in bunch crossing windows when no collisions should be present.}
\end{marginnote}

Interactions with detector material can cause an $R$-hadron to lose momentum and even come to a stop, provided it lives long enough. In scenarios with very large lifetimes, at the 10 ns scale and beyond, searches have been performed for $R$-hadrons getting stuck in material and decaying at a much later time in time windows when no collision is present and no signals are expected in the detector~\cite{ATLAS:2021mdj,CMS:2017kku}. Such searches require yet another level of custom simulation treatment, and estimating the acceptance in such an analysis requires detailed analysis of the bunch scheme, fill schedule, and run schedule. While most searches are sensitive to the integrated luminosity, stopped particle searches require high integrated luminosity per unit time where the relevant timescale is a function of the hypothesized particle lifetime. Such searches can set constraints on $R$-hadrons with lifetimes on the years scale, but they tend to be less sensitive than their direct detection counterparts since a sizable fraction of $R$-hadrons are likely to be charged. If there were some mechanism for neutral states to dominate, stopped particle searches may have unique sensitivity, but since the hadronization mechanism is SM QCD, this is unlikely. More optimistically, in the case of a discovery, stopped particle searches may be uniquely suited to measure the lifetime of such a particle, if it is greater than $O(10)$~ns.

\section{SENSITIVITY BEYOND COLLIDERS}

In this review, we discuss searches for direct production of sparticles at colliders, with a focus on rare and challenging SUSY signatures at the LHC. Searches for direct production of sparticles are complemented by searches for indirect evidence of BSM physics via precision measurements sensitive to loop effects and by searches for cosmologically-produced dark matter via either direct or indirect searches. While a comprehensive discussion of these measurements, searches, and their implications for SUSY is beyond the scope of this review, here we provide a few pointers to complementary constraints on supersymmetric models.

If SUSY solves the DM puzzle, there should be neutralinos (or $\tilde{G}$ or axinos) continuously zipping around and through us, much as we are spritzed with cosmic microwave background photons and cosmologically-produced neutrinos. Therefore, direct detection searches for WIMP DM, including the XENON1T~\cite{XENON:2018voc} and LUX-ZEPLIN~\cite{LZ:2022lsv} experiments, place powerful constraints on SUSY, if the LSP is neutral, stable, at the weak-scale, and composes a sizable fraction of the DM. The interactions of the DM with the detectors depend on the DM spin, mass, and couplings, and therefore implications for SUSY depend on the LSP properties. For a recent review of the implications of the null results of the ton-scale noble liquid WIMP searches on weak-scale SUSY, see Ref.~\cite{theoristSUSY_review}.

The most general SUSY models introduce a plethora of CP-violating couplings. While these can help address the observed matter anti-matter asymmetry in small doses, they can quickly lead to predictions in tension with experimental measurements sensitive to CP-violating terms. These measurements include CP-violation in the Kaon sector, such as the NA48 and KTeV experiments~\cite{NA48:1999szy, KTeV:2008nqz}, and measurements of the electron, neutron, and mercury Electric Dipole Moments, including recent results from the ACME collaboration~\cite{ACME:2018yjb}. For recent overviews of the implications of these measurements on SUSY models, see Refs.~\cite{annurev-nucl-102419-054905,ParticleDataGroup:2024cfk,Nakai:2016atk}.

Generic SUSY models introduce new sources of flavor violation in the soft SUSY-breaking sector, leading to flavor-changing neutral currents and rare decays at rates well above Standard Model predictions~\cite{Fok:2010vk}. Experiments searching for flavor violation therefore constrain SUSY parameters, including searches for flavor-violating decays of the muon, such as MEG~\cite{MEG:2016leq} and the upcoming Mu2E experiment~\cite{Bernstein:2019fyh}, rare Kaon decays~\cite{Nakai:2016atk}, and precise measurements of b-quark mixing parameters as well as searches for rare $B$-meson decays~\cite{Altmannshofer:2009ne,LHCb:2018roe}. Additionally, SUSY can generate significant contributions to the anomalous magnetic moment of the muon via loops involving sleptons, charginos, and neutralinos, a possibility highlighted by the recently-resolved tension between measurements and theoretical calculations of the muon magnetic moment~\cite{Muong-2:2025xyk,Endo:2021zal, Aliberti:2025beg}.

\section{CONCLUSIONS}

The search for supersymmetry has played a central role in the pursuit of BSM physics at colliders, up to and including the LHC. Regardless of how strong one's SUSY priors may be, supersymmetry provides a comprehensive and internally consistent framework for generating collider signatures, making it invaluable for developing generic search strategies that can be straightforwardly mapped onto other BSM scenarios. Moreover, specific SUSY models have motivated searches for rare and/or challenging signatures that might otherwise have been overlooked, yet which remain well-motivated both within SUSY and in broader BSM contexts. The program of SUSY searches, particularly within the CMS and ATLAS collaborations, has achieved remarkable sensitivity to a wide range of signatures, driven in part by the large number of parameters in the MSSM and its extensions.

Low-mass, extroverted SUSY signatures were the first major targets of the program and were quickly excluded, making way for a blooming program of searches for more challenging signatures. We have attempted to outline the challenges faced by these searches, to highlight the innovative and difficult work performed by the experimental and phenomenology communities to address the challenges, and to identify some future discovery opportunities that remain. Creative advances will continue to push our sensitivity to challenging signatures, and the large dataset and higher center-of-mass energy of the HL-LHC will provide essential boosts for sensitivity to rare signatures and plenty of opportunities for discovery~\cite{CidVidal:2018eel}. Potential future colliders could open up even further discovery opportunities. 

In the design of a search program, experiments work along a spectrum of organizational philosophies, from a focus on experimental signatures (a ``signature-driven'' approach) to using model parameters to guide the program (a ``model-driven'' approach). Throughout the history of the LHC, the SUSY program has moved around this spectrum: the full model approach of early Run-1, the subsequent simplified model approach developed fully in Run 2, pMSSM scans~\cite{ATLAS:2015wrn,ATLAS:2024qmx} to identify coverage gaps, and de-simplified models studying the interplay between multiple signatures to ensure robust coverage of important parameter space. The central guiding framework of supersymmetry has allowed for a focused program to grow with the evolving experimental constraints, the cultural shifts in the experiments, and advances in theory and phenomenology. 

Despite a long history of SUSY exclusions, we haven't yet ruled out all the interesting and accessible parameter space. Note that it took us nearly 50 years to find the Higgs Boson, and that search had only one free parameter $m_H$. It perhaps shouldn't be too discouraging that it hasn't started raining yet: we have learned a lot about meteorology, and it remains cloudy. We continue to find the search for SUSY to be both motivated and motivating, for its philosophical, aesthetic, and practical features, and we look forward to exciting developments in searches for supersymmetry in the decades to come.

\FloatBarrier

\section*{DISCLOSURE STATEMENT}
The authors are not aware of any affiliations, memberships, funding, or financial holdings that might be perceived as affecting the objectivity of this review. 

\section*{ACKNOWLEDGMENTS}
We thank Howard Baer, Mike Hance, Andreas Hoecker, and Johnny Lawless for invaluable feedback on a draft of this review. We thank Tova Holmes, Kiley Kennedy, Daniela Koeck, Jayson Paulose, Ben Rosser, and Brian Shuve for useful discussions during the preparation of this work. We thank the LHC and our colleagues in the ATLAS and CMS Collaborations for building and maintaining a vibrant search program for SUSY, and for all they have taught us about searching for SUSY.
LJ is supported by the Department of Energy, Office of Science, under Grant No. DE-SC0017996. LL is supported by the Department of Energy, Office of Science, under Grant No. DE-SC0020267 and the National Science Foundation, under Award No. 2235028.
This work was performed in part at the Aspen Center for Physics, which is supported by National Science Foundation grant PHY-2210452.


\bibliographystyle{ar-style5.bst} 
\bibliography{elusivesusy}

\FloatBarrier
\section*{SUPPLEMENTARY MATERIAL}

\begin{figure}
\centering
    \includegraphics[width=4.5in]{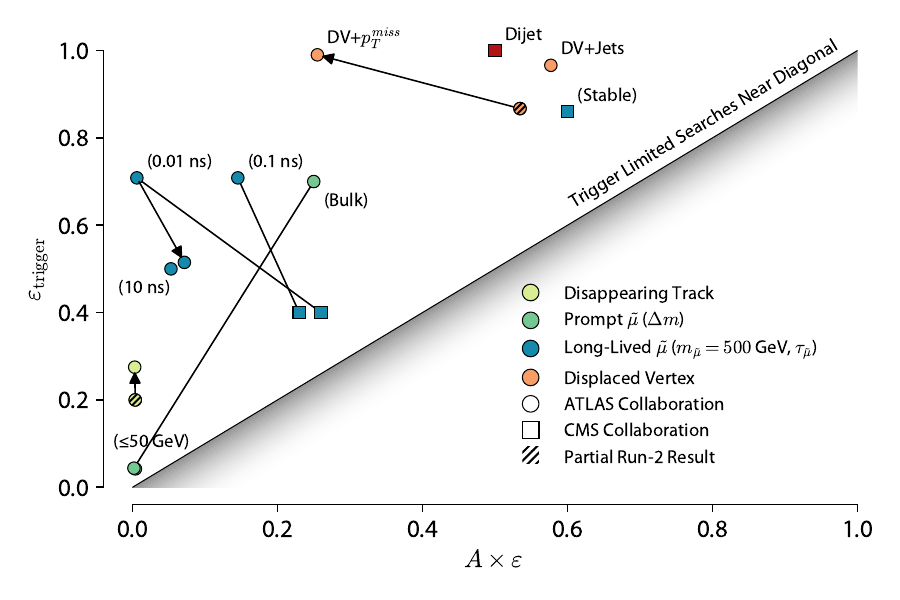}
\caption{
The contribution to the $A\times\varepsilon$ due to the trigger efficiency factor ($\varepsilon_{\textrm{trigger}}$), which is at most equal to $\varepsilon$, is shown for a collection of Run-2 analyses. Lines represent similar signatures or model parameters, and arrows denote the evolution of an analysis.
~\cite{CMS:2019gwf,ATLAS:2023oti,ATLAS:2017tny,dvmetthesis,CMS:2024nhn,CMS:2021kdm,ATLAS:2020wjh,ATLAS:2023ios,ATLAS:2019lff,ATLAS:2025evx,ATLAS:2017oal,ATLAS:2022rme}
}
\label{fig:axe_plot_trig}
\end{figure}

\end{document}